\documentclass[prd,aps,nofootinbib,preprintnumbers]{revtex4}
\usepackage{amsmath,amssymb,epsf}
\usepackage{graphicx, pstricks}
\def\be{\begin{equation}}
\def\ee{\end{equation}}
\def\ba{\begin{equation}}
\def\ba{\begin{eqnarray}}
\def\ea{\end{eqnarray}}

\def\bq{\begin{quote}}
\def\eq{\end{quote}}

\def\Mp{M_{\mathrm{Pl}}}

\begin{document}


\title{Theory and Numerics of Gravitational Waves from Preheating after Inflation.}
\date{\today}
\author{Jean-Fran\c{c}ois Dufaux$^{1}$, Amanda Bergman$^{2}$, Gary Felder$^{2}$, Lev Kofman$^{1}$ and Jean-Philippe Uzan$^3$}
\affiliation{$^{1}$ CITA, University of
Toronto, 60 St. George Street, Toronto, ON M5S 3H8, Canada\\
$^{2}$ Department of Physics, Clark Science Center, 
Smith College Northampton, MA 01063, USA\\
$^{3}$ Institut d'Astrophysique de Paris,
Universit\'e Pierre~\&~Marie Curie - Paris VI,
CNRS-UMR 7095, 98 bis, Bd Arago, 75014 Paris, France}


\begin{abstract}
Preheating after inflation involves large, time-dependent field
inhomogeneities, which act as a classical source of gravitational
radiation. The resulting spectrum might be probed by direct detection
experiments if inflation occurs at a low enough energy scale. In this
paper, we develop a theory and algorithm to calculate, analytically
and numerically, the spectrum of energy density in gravitational waves
produced from an inhomogeneous background of stochastic scalar fields
in an expanding universe. We derive some generic analytical results
for the emission of gravity waves by stochastic media of random
fields, which can test the validity/accuracy of numerical
calculations. We contrast our method with other numerical methods in
the literature, and then we apply it to preheating after chaotic
inflation. In this case, we are able to check analytically our
numerical results, which differ significantly from previous works. We
discuss how the gravity wave spectrum builds up with time and find
that the amplitude and the frequency of its peak depend in a
relatively simple way on the characteristic spatial scale amplified
during preheating. We then estimate the peak frequency and amplitude of 
the spectrum produced in two models of preheating after hybrid inflation, 
which for some parameters may be relevant for gravity wave interferometric
experiments.
\end{abstract}
\maketitle


\section{Introduction}


An expanding Universe with matter, radiation and dark energy is practically
transparent for gravitational waves which, once produced, propagate freely 
to us. Cosmological backgrounds of gravity waves (see \cite{Grishchuk:1974ny} 
for early work, and \cite{allen}, \cite{maggiore}, \cite{hogan} for reviews) 
may thus carry unique and 
``clean'' information about the universe at very early times / high 
energies. In particular, gravity waves play an important role in the 
context of inflationary cosmology. During inflation, tensor modes are 
produced from the amplification of initial quantum fluctuations into 
classical perturbations outside the Hubble radius, due to the accelerated 
expansion of the universe \cite{starob79}. The resulting spectrum extends 
over a very large frequency range and its amplitude depends directly on 
the energy scale during inflation. These gravitational waves lead in 
particular to the B-mode polarization of the CMB anisotropy fluctuations. 
In models of inflation with large energy scales 
$\sim \left(10^{15}\,\mathrm{GeV}\right)^4$ (like chaotic inflation models), 
this imprint on the CMB anisotropies should be detectable by forthcoming 
CMB polarization experiments. On the other hand, if inflation occurs at 
lower energies (as in many hybrid inflation models), the amplitude of 
the resulting gravity waves would be too weak to be observed. 

However, there is another channel of gravity wave production related
to inflation that is not associated with super-Hubble amplification of
initial quantum fluctuations, but instead with the classical emission
of sub-Hubble gravity waves from energy sources involving large,
time-dependent inhomogeneities.  Indeed, at the end of inflation, the
inflaton decays and reheats the universe. In most models, the first
stage of this process - preheating - is dominated by an explosive and
non-perturbative production of highly inhomogeneous, non-thermal
fluctuations of the inflaton and other bosonic fields coupled to
it. In chaotic inflation models, the inflaton decays via parametric
resonant particle creation \cite{KLS}, accompanied by violent dynamics
of non-linear inhomogeneous structures of the (scalar) fields
\cite{bubbles}. In hybrid inflation models, the inflaton decays into
inhomogeneous structures via tachyonic preheating \cite{tach1},
\cite{tach2}, \cite{tach3}.  The subsequent dynamics is characterized
by turbulent interactions between Bose (scalar) waves, before the
system ultimately settles into thermal equilibrium. The inhomogeneous
decay of the inflaton, the subsequent turbulent stage and even the
thermal state are inevitably accompanied by the emission of
gravitational waves. The theory of gravity wave production from random
media of inhomogeneous scalar field(s) after inflation has to address
the question of the amplitude and typical frequencies of the resulting
stochastic background. For low energy scale inflationary models, the
frequencies of the gravity waves produced after inflation may well
occur in the range which in principle can be detected by direct
detection experiments (like LIGO/VIRGO and BBO/DECIGO). This would 
provide us with a channel for testing inflation that can complement 
the CMB data, and with a unique observational window into the subsequent 
dynamics of the very early universe.

There is no ready form of the theory of gravitational wave emission
from random media that could be immediately applied for preheating.
The formalism in Weinberg's textbook \cite{weinberg}, based on the
wave-zone approximation for localized sources in Minkowski space-time,
is widely used. The energy in gravity waves is then expressed in terms
of the double Fourier transform (in both space and time) of the
stress-energy tensor. In principle, this formalism is applicable only
for isolated sources, without expansion of the universe, and it does
not allow one to follow the evolution with time of gravity wave
emission.  There are several papers on the production of gravitational
waves from a cosmological phase transition (see \cite{Grojean:2006bp}
and references therein), due to the collision of bubbles and to the
resulting dynamics.  Gravity waves emitted from bubble collisions
have been studied in \cite{Kosowsky:1991ua}, \cite{Kosowsky:1992vn},
\cite{Kamionkowski:1993fg}, using the formalsim of \cite{weinberg} for
the process of an instantaneous (i.e. fast compared to the expansion
rate of the universe) phase transition.  Calculations of gravity waves
from hydrodynamical turbulence \cite{Kosowsky:2001xp},
\cite{Caprini:2006jb}, departed from the formalism of \cite{weinberg}
and solved instead for the wave equation of the Fourier tensor modes
sourced by the transverse-traceless part of the stress-energy
tensor. The considerations were restricted to the case of
hydrodynamical incompressible fluids, and again for an instantaneous
phase transition.

The emission of gravitational waves from preheating was addressed in the 
original paper \cite{pregw1}, and more recently in \cite{pregw2}, 
\cite{pregw3} and \cite{pregw4}. Refs.~\cite{pregw1} and \cite{pregw2} 
performed numerical calculations of the gravity waves produced from 
preheating after chaotic inflation, using the formalism of \cite{weinberg}. 
For chaotic inflation models, the spectrum of gravity wave energy density 
has a bump at very high ferquencies, of order $10^8\,\mathrm{Hz}$, which 
is not likely to be observable. In \cite{bubbles}, two of us observed that the inflaton 
fragmentation in preheating after chaotic inflation occurs through the 
formation of non-linear bubble-like field inhomogeneities. It was 
conjectured in \cite{bubbles} that the main contribution to gravity waves
emission comes from these bubble-like structures, both in the resonant and the 
tachyonic cases. Their relevance for gravity wave production in the case of 
tachyonic preheating was also pointed out in \cite{tach3}. 
These bubbles have nothing to do with phase transitions. 
They originate as non-linear structures from the realization of the 
Gaussian random field of the vacuum fluctuations. If $R$ is the 
charactersitc bubble size and $H$ is the Hubble parameter when they form, 
then the fraction of energy density in gravity waves at the time of 
production is $\rho_{\mathrm{gw}} / \rho_{\mathrm{tot}} \sim (RH)^2$ 
at frequencies $f \sim 1/R$. Ref.~\cite{pregw3} studied numerically the 
production of gravity waves from parametric resonance at different energy 
scales, solving for the equations of the metric perturbations in Fourier 
space. 
Finally, Ref. \cite{pregw4} introduced another numerical method to study 
gravity waves from tachyonic preheating after hybrid inflation. In particular, 
they displayed a case which, although involving very small coupling
constants, they concluded could be detectable by BBO. 

In this paper we develop a theoretical framework for calculating 
systematically the emission of classical gravitational waves from 
random media of dynamical scalar fields in an expanding universe.
The field inhomogeneities have very large (non-linear) amplitudes 
at scales smaller than the Hubble radius $1/H$ but the media is homogeneous 
at larger scales. Our purpose is to extract observables like the present day 
energy density of gravity waves through statistical averages over the random 
scalar fields. We derive some general results for gravity waves produced 
by stochastic media of scalar fields, which are applicable for preheating, 
phase transitions and other instances of gravity wave emission by Bose fields.
These general results will serve as tests for numerical methods suggested in 
earlier papers and as a check of our numerical results.

We have used our formalism to develop a numerical algorithm for calculating 
gravity wave production. The field dynamics are calculated with lattice 
simulations, which are then used to find the evolution of the spectrum of 
gravity waves. These spectra are then converted into physical variables 
observed today. Our formalism gives significantly different results from 
others in the literature. Therefore we pay special attention to checking 
the results by analytical calculations.

We illustrate our method with a model of preheating after chaotic inflation. 
This model provides a simple test-case for our formalism, allows easy 
comparisons with previous predictions, and allows us to verify our 
simulation results with detailed analytical calculations valid during 
the linear stage of preheating.

The production of gravity waves from preheating after hybrid inflation
is observationally more promising. We provide preliminary estimates for 
this case in this paper and will return to the issue in more detail 
in a subsequent publication.

The paper is organized as follows. In section \ref{gwprod} we develop the 
basic equations expressing gravity waves in terms of metric perturbations 
and relating these perturbations to a background of inhomogeneous scalar 
fields in an expanding universe. We also identify typical scalar field 
configurations that do and do not emit gravitational radiation. 
In section \ref{energyspectrum} we compute the spectrum of energy density 
in gravity waves emitted by stochastic media, which includes taking 
spatial averages and ensemble averages over different realizations of the 
random fields. The spectra are then converted into present-day physical 
variables. In section \ref{Analytic}, we outline analytical calculations of 
gravitational wave emission from different stages of preheating and 
thermalization, and we illustrate the formalism developed in the previous 
sections. We present our numerical method in section \ref{Numerical}, and 
we contrast it with previous numerical calculations of gravity waves from 
preheating. Section \ref{lamphi4} contains our numerical and analytical 
results for the case of preheating after $\lambda \phi^4$ inflation. We 
discuss the influence of different stages of the scalar field dynamics 
on gravity wave production, and we express the amplitude and frequency 
of the spectrum's peak in terms of the model's parameters. We then perform 
a detailed analytical check of our numerical results, part of which is
in an Appendix. In section \ref{hybrid}, we estimate the 
amplitude and frequency of gravity waves produced in two different models 
of preheating after hybrid inflation. We conclude in section 
\ref{conclusions} with a summary of our results and perspectives.


\section{Emission of Gravity Waves by Inhomogeneous Scalar Field Sources}
\label{gwprod}


In this section we present the basic equations for classical gravitational 
waves emitted by a background made of inhomogeneous scalar fields. 
We also spell out a rather obvious but very useful theorem stating that the superposition of scalar 
fields waves with a ``particle-like'' dispersion relation does not emit gravity waves.

\subsection{Equations for Gravitational Waves with Sources in an Expanding Universe}
\label{hij}

We consider several inhomogeneous scalar fields, denoted collectively by $\{ \phi_a, a = 1,2, ... \}$, 
with energy-momentum tensor 
\be
\label{Tscal}
T_{\mu \nu} = \partial_{\mu} \phi_a\,\partial_{\nu} \phi_a 
- g_{\mu \nu}\;\left(\frac{1}{2}\,g^{\rho \sigma}\,\partial_{\rho} \phi_a\,\partial_{\sigma} \phi_a + V\right)
\ee
where repeated indices $a$ are summed. During preheating, the inflaton decays inhomogeneously and some of the fields 
coupling to it are significantly amplified. The fields are rather homogeneous at large scales (at the Hubble radius 
and beyond), but highly inhomogeneous inside the Hubble radius. These field inhomogeneities at small scales cannot be 
treated as small perturbations\footnote{In fact, they correspond typically to density contrasts $\delta\rho / \rho \sim \mathrm{tens}$}. 
They participate in the evolution of the background scale factor through the average 
of the energy momentum tensor $\langle T_{\mu \nu} \rangle$ in the Einstein equations. Here we consider the linear response 
of the metric perturbation $\delta g_{\mu \nu}$ to the inhomogeneous part of $T_{\mu \nu}$. We will work at linear order 
in $\delta g_{\mu \nu}$ because its coupling to $T_{\mu \nu}$ is suppressed by the Planck mass $\Mp$, and the typical mass 
scales involved in the energy-momentum tensor are much lower than $\Mp$. Among the different components of the metric 
perturbation, gravitational waves are the only physical degrees of freedom which propagate and carry energy out of the 
source, see e.g. \cite{flanagan}. 

In a (spatially flat) Robertson-Walker background, gravitational waves may be represented by the transverse and traceless part 
of the spatial metric perturbation
\be
ds^2 = g_{\mu \nu}\,dx^{\mu} dx^{\nu} = a^2(\tau)\,\left[-d\tau^2 + \left(\delta_{ij} + h_{ij}\right)\,dx^i dx^j\right]
\ee
with\footnote{Here and in the following, Latin indices $i, j, ...$ run over the $3$ spatial coordinates, 
and repeated indices are summed. They are raised and lowered with the Kronecker symbol $\delta_{ij}$, 
so we don't distinguish between upper and lower indices.} $\partial_i h_{ij} = h_{ii} = 0$. 
The perturbation $h_{ij}$ corresponds to two independent tensor degrees of freedom 
and has the equation of motion 
\be
\label{eom}
h_{ij}'' + 2\,\frac{a'}{a}\,h_{ij}' - \mathbf{\nabla}^2 h_{ij} = 16\pi G\,a^2\,\Pi_{ij}^{\mathrm{TT}} \; ,
\ee
where a prime denotes a derivative with respect to conformal time $\tau$.

Free gravitational waves obey the equation (\ref{eom}) without the source term. They can be quantized, and one can study 
the amplification of their vacuum quantum fluctuations in an expanding universe. 
An especially important case is exponential expansion of the universe during inflation, when tensor mode quantum fluctuations 
lead to a stochastic background of classical long-wavelength gravitational waves.

We will consider a different, complementary situation, when quantum effects are negligible, and classical gravitational waves 
are generated by a non-zero source term in Eq.~(\ref{eom}). The source term $\Pi_{ij}^{\mathrm{TT}}$ is the transverse-traceless 
part ($\partial_i \Pi_{ij}^{\mathrm{TT}} = \Pi_{ii}^{\mathrm{TT}} = 0$) of the anisotropic stress $\Pi_{ij}$
\be
\label{anis}
a^2\,\Pi_{ij} = T_{ij} - \langle p \rangle \, g_{ij}
\ee
where $\langle p \rangle$ is the background homogeneous pressure. Formally, the second term in the RHS of Eq.~(\ref{anis}) and the 
second term in the $(i,j)$ components of the RHS of Eq.~(\ref{Tscal}) involve the metric perturbation $h_{ij}$, through 
$g_{ij} = \delta_{ij} + h_{ij}$. This  gives a contribution in 
$\frac{16 \pi G}{3}\,\langle \partial_k \phi_a \partial_k \phi_a \rangle \, \bar{h}_{ij}$ (where $\bar{h}_{ij}$ is defined in 
(\ref{barh})) in the RHS of Eq.~(\ref{eom2}) below. We shall not include this term since it emerges only at second order in 
the gravitational coupling and is  negligible at sub-Hubble scales. 

There are different ways to solve the wave equation (\ref{eom}). One can use Green functions in configuration space, but the 
transverse-traceless projection then involves inconvenient non-local operators. Another method~\cite{weinberg}, for the 
harmonic gauge ($\partial^{\mu} h_{\mu \nu} = 0$) in Minkowski spacetime, is based on the wave-zone approximation of the 
solution in configuration space, expressed in terms of the double Fourier transform (in both time and space) 
$T_{ij}(\mathbf{k}, \omega)$ of the stress-energy tensor. This method was used in the context of preheating in 
Refs.~\cite{pregw1}, \cite{pregw2}. Formally, it is applicable only for isolated sources, without expansion of the 
universe, and requires the knowledge of the whole evolution of $T_{ij}$ with time. 

Here we develop another formalism, which is better suited to cases such as reheating with extended sources or continuous media 
in an expanding universe, and which allows us to follow the evolution of gravity waves with time. 
We will work in spatial Fourier space, with the convention 
\be
\label{fourier}
f(\mathbf{x}) = \int \frac{d^3\mathbf{k}}{(2\pi)^{3/2}}\,f(\mathbf{k})\,e^{-i \mathbf{k} \mathbf{x}} \ .
\ee 
With the further redefinition 
\be
\label{barh}
\bar{h}_{ij} = a\,h_{ij} \;\; ,
\ee
Eq.~(\ref{eom}) gives
\be
\label{eom2}
\bar{h}_{ij}''(\mathbf{k}) + \left(k^2 - \frac{a''}{a}\right)\,\bar{h}_{ij}(\mathbf{k}) = 16 \pi G\,a^3\,\Pi_{ij}^{\mathrm{TT}}(\mathbf{k})
\ee
where $k^2 = \mathbf{k}^2$ is the square of the {\it comoving} wave-number.

Given a symmetric tensor $\Pi_{ij}$, its transverse-traceless part is easily obtained in momentum space 
(but is non-local in configuration space) by the projection (see e.g. \cite{misner})
\be
\label{TTproj}
\Pi_{ij}^{\mathrm{TT}}(\mathbf{k}) = \mathcal{O}_{ij , lm}(\mathbf{\hat{k}}) \, \Pi_{lm}(\mathbf{k}) = 
\left[P_{il}(\mathbf{\hat{k}})\,P_{jm}(\mathbf{\hat{k}}) - \frac{1}{2}\,P_{ij}(\mathbf{\hat{k}})\,P_{lm}(\mathbf{\hat{k}})\right]\,
\Pi_{lm}(\mathbf{k}) 
\ee
with
\be\label{Pij}
P_{ij}(\mathbf{\hat{k}}) = \delta_{ij} - \hat{k}_i\,\hat{k}_j
\ee
where $\mathbf{\hat{k}} = \mathbf{k} / k$ is the unit vector in the $\mathbf{k}$ direction. The operators $P_{ij}$ are projectors on the subspace 
orthogonal to $\mathbf{k}$, satisfying $P_{ij} k_i = 0$ and $P_{ij}\,P_{jl} = P_{il}$. From this it follows directly that 
$k_i\,\Pi_{ij}^{\mathrm{TT}} = \Pi_{ii}^{\mathrm{TT}} = 0$. 

In our case, the second term in the RHS of Eq.~(\ref{anis}) and the second term in the $(i,j)$ components of the RHS of Eq.~(\ref{Tscal}) 
are pure trace ($g_{ij} = \delta_{ij}$, see the discussion below Eq.~(\ref{anis})), and therefore they do not contribute to the 
transverse-traceless part. The relevant part of the energy-momentum tensor is then given by the product of the spatial derivatives 
of the fields, giving a convolution in Fourier space
\be
\label{TijTT}
a^2\,\Pi_{ij}^{\mathrm{TT}}(\mathbf{k}) = T_{ij}^{\mathrm{TT}}(\mathbf{k}) = 
\mathcal{O}_{ij , lm}(\mathbf{\hat{k}})\,\left\{\partial_l \phi_a \, \partial_m \phi_a\right\}(\mathbf{k}) 
= \mathcal{O}_{ij , lm}(\mathbf{\hat{k}})\,\int \frac{d^3\mathbf{p}}{(2\pi)^{3/2}}\,p_l\,p_m\,\phi_a(\mathbf{p})\,\phi_a(\mathbf{k} - \mathbf{p})
\ee  
where $\left\{\partial_l \phi_a \, \partial_m \phi_a\right\}(\mathbf{k})$ denotes the Fourier transform of 
$\partial_l \phi_a \, \partial_m \phi_a$. 
In the last equality, we dropped a term in $k_m$ which vanishes when contracted with $\mathcal{O}_{ij , lm}(\mathbf{\hat{k}})$. 

It should be clear that the process of gravity wave production that we consider here is very different from the one during inflation. During 
inflation, initial quantum fluctuations are amplified into super-Hubble stochastic classical perturbations, as a result of the 
accelerated expansion, that is through the term $a'' / a$ in Eq.~(\ref{eom2}). In this case, the inhomogeneous part of the scalar 
field (inflaton) corresponds to a small perturbation (compared to the homogeneous part), and the source term $\Pi_{ij}^{\mathrm{TT}}$ 
vanishes at linear\footnote{Gravitational waves can be produced classically from the second-order scalar perturbations generated during 
inflation \cite{Ananda:2006af, Osano:2006ew, Baumann:2007zm}.} 
order in $\delta \phi(\mathbf{x}, \tau)$, see Eq.~(\ref{TijTT}). This is to be contrasted to the case considered here, 
where the inhomogeneous part of the scalar fields cannot be treated as small perturbations, and $\Pi_{ij}^{\mathrm{TT}}$ act as a classical 
source. Note also that in Eq.~(\ref{eom2}), the term in $a'' / a$ vanishes when the equation of state is the one of radiation, $w = 1/3$. 
In general, this is not exactly satisfied during preheating. Indeed, when the inflaton potential is quadratic around its minimum, the average 
equation of state usually jumps during preheating from $w = 0$ to an intermediate value that is close, but not exactly equal, to 
$w = 1/3$, see \cite{eos}. However, the dominant part of the gravity wave spectrum is produced well inside the Hubble radius at the time 
of production, so that $a'' / a$ is negligible: $a'' / a \, \sim \, a^2 H^2 \ll k^2$. In principle we can keep the term in $a'' / a$ and 
solve Eq.~(\ref{eom2}) with a Green function more complicated than (\ref{h}) below. However, as we argued, this term can be dropped and 
Eq.~(\ref{eom2}) then reduces to 
\be
\label{eom3}
\bar{h}''_{ij}(\tau, \mathbf{k}) + k^2\,\bar{h}_{ij}(\tau, \mathbf{k}) = 16 \pi G\,a(\tau)\,T_{ij}^{\mathrm{TT}}(\tau, \mathbf{k})
\ee
where we have used Eq.~(\ref{TijTT}). 

The solution with $h_{ij}(\tau_i) = h_{ij}'(\tau_i) = 0$ (assuming no gravity waves at the scale $k$ before the initial time $\tau_i$) is 
given by a simple Green function
\be
\label{h}
\bar{h}_{ij}(\tau, \mathbf{k}) = \frac{16\pi G}{k}\,\int_{\tau_i}^{\tau} d\tau'\,\sin\left[k\,(\tau - \tau')\right]\,
a(\tau')\,T_{ij}^{\mathrm{TT}}(\tau', \mathbf{k})  
\ee
for $k \neq 0$. Note that
\be
\label{h'}
\bar{h}'_{ij} = 16 \pi G\,\int_{\tau_i}^{\tau} d\tau'\,\cos\left[k\,(\tau - \tau')\right]\,a(\tau')\,T_{ij}^{\mathrm{TT}}(\tau', \mathbf{k}) \ .
\ee

If the source eventually becomes negligible after some time $\tau = \tau_f$ (see below), the waves then freely propagate. 
The modes were sub-Hubble at the time of production and remain sub-Hubble until today. The corresponding solution of 
Eq.~(\ref{eom3}) without a source is simply
\be
\label{hfree}
\bar{h}_{ij}(\tau, \mathbf{k}) = A_{ij}(\mathbf{k})\,\sin\left[k (\tau - \tau_f)\right] + 
B_{ij}(\mathbf{k})\,\cos\left[k (\tau - \tau_f)\right] \;\;\;\; \mbox{ for } \;\;\; \tau \geq \tau_f \ .
\ee
Matching $h_{ij}$ and $h_{ij}'$ at $\tau = \tau_f$ with (\ref{h}, \ref{h'}) gives
\ba
\label{AB}
A_{ij}(\mathbf{k}) = \frac{16\pi G}{k}\,\int_{\tau_i}^{\tau_f} d\tau'\,\cos\left[k\,(\tau_f - \tau')\right]\,
a(\tau')\,T_{ij}^{\mathrm{TT}}(\tau', \mathbf{k})
\nonumber \\
B_{ij}(\mathbf{k}) = \frac{16\pi G}{k}\,\int_{\tau_i}^{\tau_f} d\tau'\,\sin\left[k\,(\tau_f - \tau')\right]\,
a(\tau')\,T_{ij}^{\mathrm{TT}}(\tau', \mathbf{k}) \ .
\ea

Before we develop our formalism further, we will discuss the emission
of gravity waves by a special but interesting scalar field 
configuration.

\subsection{No-go Theorem: No Gravity Waves from Scalar Field Waves}
\label{nogw}

For the amplitude of the gravity wave, we obtained above
\be
\label{hrec}
\bar{h}_{ij}(\tau, \mathbf{k}) = \frac{16\pi G}{k}\, \mathcal{O}_{ij , lm}(\mathbf{\hat{k}})\,
\int_{\tau_i}^{\tau} d\tau'\,\sin\left[k\,(\tau - \tau')\right]\,
a(\tau') \, \int \frac{d^3\mathbf{p}}{(2\pi)^{3/2}}\,p_l\,p_m\,\phi_a(\tau', \mathbf{p})\,\phi_a(\tau', \mathbf{k} - \mathbf{p}) \ .
\ee

It is instructive to consider the very simple example of gravitational waves emitted by a medium made of a real free massive scalar field 
$\phi$ obeying the Klein-Gordon equation in Minkowski spacetime ($a(\tau) = 1$ in the equations above). In this case, the Fourier transform 
of the field reads
\be
\phi(\mathbf{p}, \tau) e^{i  \mathbf{p} \mathbf{x} } =  b(\mathbf{p}) \, e^{\pm i \omega_p \tau +i \mathbf{p} \mathbf{x}} 
\ee
with the dispersion relation 
\be
\label{dispersion}
\omega_p^2 = p^2 + m^2 \, ,
\ee
where $m$ is the mass of $\phi$. Decomposing the circular functions in Eq.~(\ref{hrec}) or (\ref{AB}) into 
exponentials, and reordering the integrals over $d\tau$ and $d^3\mathbf{p}$ we then have 
\be
\label{staphase}
h_{ij}(\tau, \mathbf{k}) \propto e^{\pm i k \tau} \mathcal{O}_{ij , lm}(\mathbf{\hat{k}})\,\int d^3\mathbf{p}\,p_l\,p_m\,  
b(\mathbf{p}) b(\mathbf{k} - \mathbf{p}) \int_{\tau_i}^{\tau} d\tau'\,
e^{i\,(\pm \, \omega_p \, \pm \, \omega_{|\mathbf{k} - \mathbf{p}|} \, \pm \, k) \, \tau'} \ .
\ee

\begin{figure}[htb]
\centering \leavevmode \epsfxsize=5cm
\epsfbox{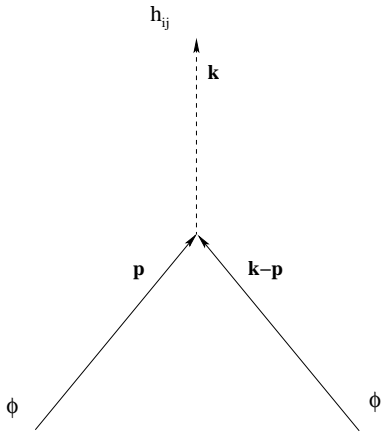}
\caption{Would be emission of a graviton $h_{ij}$ with momentum $\mathbf{k}$ from the annihilation of two scalar waves $\phi(\mathbf{p})$
and $\phi(\mathbf{k} - \mathbf{p})$ with momenta $\mathbf{p}$ and $\mathbf{k} - \mathbf{p}$. Helicity $2$ of the emitted graviton 
cannot match the helicity zero of the incoming scalar waves.}
\label{diag}
\end{figure} 

Eq.~(\ref{staphase}) corresponds to trilinear interactions between two
$\phi$-particles and one graviton, with different signs 
in the phase inside the time integral corresponding to different channels of interaction. In the limit of large time $\tau$ compared 
to the frequencies of the particles, the time integrals reduce to Dirac delta functions enforcing energy conservation, such as 
$\omega_p +\omega_{|\mathbf{k} - \mathbf{p}|} = k$. Such trilinear interactions can be sketched by the diagram shown in the 
Figure~\ref{diag}. Momentum conservation is encoded in the convolution $\phi(\tau', \mathbf{p})\,\phi(\tau', \mathbf{k} - \mathbf{p})$.
Energy and momentum conservations may be satisfied only for massless $\phi$-particles, and only with $\mathbf{k}$ parallel to $\mathbf{p}$. 
However, when $\mathbf{k}$ is parallel to $\mathbf{p}$, the projector operator brings the gravity wave amplitude to zero, 
$\mathcal{O}_{ij , lm}(\mathbf{\hat{k}})\,p_l\,p_m = 0$ in Eq.~(\ref{staphase}). The reason for this is the conservation of helicity, 
which forbids interactions between free scalar waves and a graviton at linear order in the gravitational coupling. Interactions involving 
several gravitons are possible but further suppressed by the Planck mass. If instead of scalar waves the source of gravitational radiation 
is a superposition of vector field waves (photons or gauge bosons), gravitons can be emitted already at first order in the gravitational 
coupling, since diagrams like in Figure~\ref{diag} but with vector fields carrying helicity 1 respect helicity conservation. For example, a
thermal bath of photons emits gravitational waves \cite{ford}, while a thermal bath of massless scalars does not (see sub-section 
\ref{fourth} below). 

Let us try to understand how far this no-go result can be extended, and which configurations of the scalar fields can lead to gravity 
wave emission. Suppose that instead of free scalar waves we deal with
interacting scalar waves (which is typical for preheating). 
Then, instead of (\ref{dispersion}), the dispersion relation of the field $\phi$ will involve the other fields interacting with it. For 
instance, for a quartic interaction $g^2\,\phi^2\,\psi^2$ between $\phi$ and another scalar field $\psi$, we have 
\be
\label{dispersion2}
\omega_p^2 = p^2 + m^2 +  g^2 \psi^2   \ .
\ee
If the frequencies $\omega_p$ vary adiabatically with time, the time evolution of $\phi$ can be described by the WKB approximation, 
$\phi(\mathbf{p}, \tau) \propto e^{\int^{\tau} d\tau' \omega_p}$ in 
Eqs. (\ref{hrec}), (\ref{staphase}). Then
trilinear interactions again satisfy energy conservation 
(technically, this results in this case from the stationary phase approximation), and the no-go result is intact. Therefore, we can 
formulate the no-go theorem:\\
{\it Scalar field configurations which can be represented as the superposition of waves with wave-like dispersion relations
and adiabatically varying frequencies do not emit gravity waves at first order in the gravitational coupling}. 

This result covers several interesting cases. For instance, during preheating after chaotic inflation, scalar field waves are 
produced only during short intervals of time and vary adiabatically between those instances. Therefore, we expect no gravity 
wave emission during the intermediate adiabatic regimes. We will treat this case in detail below, in sub-section \ref{analytics}.  
Between preheating and thermalisation the scalar fields enter a stage of Kolmogorov turbulence, characterized by the irreversible 
dynamics of weakly interacting scalar waves. We do not expect gravity waves to be emitted from this stage neither. Therefore, in 
the context of preheating, the time $\tau = \tau_f$ introduced in the previous sub-section in Eq.~(\ref{hfree}), should be formally 
associated with the onset of the regime of weak turbulence. Another interesting case is a thermal bath of interacting scalar fields, 
which also does not emit gravity waves. We will discuss these cases in more details in Section \ref{Analytic}. 

Now, let us try to understand which configurations of scalar fields
{\it can} lead to the emission of gravitational waves?
The no-go theorem is violated when the dispersion relation is not wave-like, and when the frequency does not vary adiabatically. 
We already mentioned one example of non-adiabatic change of the frequency of the field which is parametrically produced during preheating. 
In this case gravity waves are emitted even by massive scalars. Another example is the bubble-like configuration described by non-linear 
scalar fields where the wave-like dispersion is violated, as for bubbles formed from a first order phase transition.  
The collisions of scalar field bubbles emit gravity waves.


\section{GW Energy Density Spectrum}
\label{energyspectrum}


In this section, we derive the spectrum of energy density in gravitational waves emitted by the random media of the scalar fields. 
This involves taking the spatial average of bilinear combinations of the (transverse-traceless) metric perturbation. This is  what 
we will use in our numerical simulations. We develop this approach in sub-section \ref{spatav}. In sub-section \ref{weinberg} we 
relate our formula to the Weinberg formula for $\frac{d E_{\mathrm{gw}}}{d\Omega}$, which is often used in the literature.
However, the simulations rely on a particular realization of the initial quantum fluctuations for the scalar fields. 
Analytical calculations can be advanced further by taking  average over the ensemble of different realizations, and we calculate 
the gravity waves spectrum in this way in sub-section \ref{ensav}. Finally, in sub-section \ref{speto}, we present the rescalings 
needed to convert the spectrum into the present-day physical variables.

\subsection{GW Energy Density}
\label{spatav}

The energy density carried by a gravity wave cannot be localized in regions smaller than its  wavelength, but  
it can be defined as an average over a volume $V$ of several wavelengths' size (see e.g. \cite{misner})
\be
\label{rhogwmtw}
\rho_{\mathrm{gw}} = \frac{1}{32 \pi G}\,\langle \dot{h}_{ij}(t, \mathbf{x})\,\dot{h}_{ij}(t, \mathbf{x}) \rangle_V
\ee 
where a dot denotes a derivative with respect to cosmic time $t = \int a\,d\tau$. 
In terms of conformal time and $\bar{h}_{ij}$ in (\ref{barh}), we have
\be
\label{rhohbar}
\dot{h}_{ij}\,\dot{h}_{ij} = \frac{1}{a^4}\,\left(\bar{h}_{ij}'\,\bar{h}_{ij}' + 2 a H\,\bar{h}_{ij}\,\bar{h}_{ij}' 
+ a^2 H^2\,\bar{h}_{ij}\,\bar{h}_{ij}\right)
\ee
For sub-Hubble wavelengths, $k / a \gg H$, the second and third terms are negligible with respect to the first one. We therefore have
\be
\label{rhogw}
\rho_{\mathrm{gw}} = \frac{1}{32 \pi G a^4}\,\langle \bar{h}_{ij}'(\tau, \mathbf{x})\,\bar{h}_{ij}'(\tau, \mathbf{x}) \rangle_V = 
\frac{1}{32 \pi G a^4}\,\frac{1}{V}\,\int d^3\mathbf{k}\,\bar{h}_{ij}'(\tau, \mathbf{k})\,\bar{h}_{ij}'^{*}(\tau, \mathbf{k}) 
\ee
where $*$ denotes the complex conjugate. In the second equality, we expanded each 
$\bar{h}_{ij}(\tau, \mathbf{x})$ into Fourier components $\bar{h}_{ij}(\tau, \mathbf{k})$, and then calculated the remaining spatial average as 
\be
\frac{1}{V}\,\int_{V \gg \lambda^3} d^3\mathbf{x}\,e^{-i (\mathbf{k} + \mathbf{k}') \mathbf{x}} = 
\frac{(2\pi)^3}{V}\,\delta^{(3)}(\mathbf{k} + \mathbf{k}')  
\ee
where the (comoving) volume $V$ has large dimensions compared to the (comoving) wavelengths $\lambda$. The volume factor appears for dimensional 
reasons due to our use of a continuous Fourier transform
(\ref{fourier}), as opposed to a Fourier series. In the lattice simulations, $V$ will 
correspond to the volume of the box in configuration space. The final results are independent of $V$.

For $h_{ij}(\tau, \mathbf{k})$ in Eq.~(\ref{rhogw}), we use (\ref{hfree}) corresponding to the free waves propagating up to now after the 
emission process is completed. Suppose that today we are not interested in the resolution of the oscillation of $\bar{h}_{ij}(\tau, \mathbf{k})$    
with time, so we average over a complete period of oscillation $T = \frac{2\pi}{k}$
\ba
\label{timeav}
&& \langle \bar{h}_{ij}'(\tau, \mathbf{k})\,\bar{h}_{ij}'^{*}(\tau, \mathbf{k}) \rangle_{\mathrm{T}} 
= \frac{k^2}{2}\,\sum_{i,j} \left(|A_{ij}|^2 + |B_{ij}|^2\right) = \nonumber \\ && 
\frac{(16\pi G)^2}{2}\,\sum_{i,j} \left\{ \left| \int_{\tau_i}^{\tau_f} d\tau'\,\cos\left[k\,(\tau_f - \tau')\right]\,
a(\tau')\,T_{ij}^{\mathrm{TT}}(\tau', \mathbf{k}) \right|^2 + 
\left| \int_{\tau_i}^{\tau_f} d\tau'\,\sin\left[k\,(\tau_f - \tau')\right]\,a(\tau')\,T_{ij}^{\mathrm{TT}}(\tau', \mathbf{k}) \right|^2 \right\}
\ea
where $\sum_{i,j} \left| X_{ij}\right|^2 = X_{ij}\,X_{ij}^*$, and we used (\ref{AB}) in the last equality.  
Expanding the cosine and sine above in factors of $\cos(k \tau_f)$ and $\sin(k \tau_f)$, these factors go out 
of the integrals and eventually give $\cos^2(k \tau_f) + \sin^2(k \tau_f) = 1$. Plugging the result into (\ref{rhogw}), we get
\be
\label{rhof}
\rho_{\mathrm{gw}} = \frac{4\pi G}{a^4}\,\frac{1}{V}\,\int d^3\mathbf{k}\,\sum_{i,j} 
\left\{ \left| \int_{\tau_i}^{\tau_f} d\tau'\,\cos\left(k\,\tau'\right)\,a(\tau')\,T_{ij}^{\mathrm{TT}}(\tau', \mathbf{k}) \right|^2 + 
\left| \int_{\tau_i}^{\tau_f} d\tau'\,\sin\left(k\,\tau'\right)\,a(\tau')\,T_{ij}^{\mathrm{TT}}(\tau', \mathbf{k}) \right|^2 \right\}
\ee

From (\ref{rhof}), we construct  the spectrum of gravity wave energy density per unit logarithmic frequency interval 
\be
\label{defSk}
\left(\frac{d \rho_{\mathrm{gw}}}{d \ln k}\right)_{\tau > \tau_f} = \, \frac{S_k(\tau_f)}{a^4(\tau)}
\ee
where we have defined
\be
\label{Xk}
S_k(\tau_f) = \frac{4\pi G \, k^3}{V}\,\int d\Omega\,\sum_{i,j} 
\left\{ \left| \int_{\tau_i}^{\tau_f} d\tau'\,\cos\left(k\,\tau'\right)\,a(\tau')\,T_{ij}^{\mathrm{TT}}(\tau', \mathbf{k}) \right|^2 + 
\left| \int_{\tau_i}^{\tau_f} d\tau'\,\sin\left(k\,\tau'\right)\,a(\tau')\,T_{ij}^{\mathrm{TT}}(\tau', \mathbf{k}) \right|^2 \right\} \ .
\ee
This is the main quantity that we will compute numerically. It does not depend on the subsequent cosmological evolution, which dilutes 
the energy density in gravity waves and redshifts their frequencies. We will take care of the corresponding rescalings below in 
sub-section \ref{speto}. The final spectrum today is then obtained from (\ref{Xk}) where $\tau_f$ is the time when the source becomes 
negligible. However, we will also study how the gravity waves spectrum evolves with the time $\tau_f$ before the source becomes negligible, 
in order to investigate the influence of the different stages of the scalar fields dynamics on the production of gravity waves.
 
It is interesting to analyze Eq.~(\ref{Xk}) in the quadrupole approximation for large wavelengths (small k). In this approximation, 
one may drop the $k$-dependence in the source term $T_{ij}^{\mathrm{TT}}(\tau', \mathbf{k})$. We may then distinguish two different 
regimes where the spectrum of the energy density in gravity waves depends in a simple way on the frequency. For sufficiently small 
$k$, the sine and cosine in the time integrals of Eq.~(\ref{Xk}) vary more slowly with time than the source and can be taken out 
of the integrals. In this case, the $k$-dependence of $S_k$ comes only from the pre-factor in $k^3$, so the gravity wave 
spectrum varies as the cube of the frequency. On the other hand, for larger $k$, there may be an intermediate regime where the 
quadrupole approximation is still valid but the sine and cosine vary more rapidly with time than the source. In this case, 
the time integrals of the factors in $\cos\left(k\,\tau'\right)$ and $\sin\left(k\,\tau'\right)$ in Eq.~(\ref{Xk}) are proportional 
to $1/k$. Overall, this gives $S_k \propto k$, so the gravity waves spectrum varies linearly with the frequency.

\subsection{Relation to the Weinberg Formula for $\frac{d E_{\mathrm{gw}}}{d\Omega}$}
\label{weinberg}

The formalism developed above is designed for cases, such as reheating, where the source or continuous media extends over 
the whole expanding universe, and is active - with respect to gravity wave emission - for a limited period of time 
(until $\tau_f$). It may be directly extended to systems emitting gravity waves continuously - i.e. active all the time - 
by taking the limit $\tau_f \rightarrow \infty$. In this case, we may relate our formalism to the one derived by 
Weinberg~\cite{weinberg} for isolated sources in the wave-zone limit (i.e. at distances large compared to the wavelengths and to 
the size of the localized source) in Minkowski spacetime, which is often used in the literature.

Consider the double Fourier transform of the energy-momentum tensor in both space and time (in Minkowski spacetime)
with the conventions of ~\cite{weinberg}
\be
\label{double}
T_{ij}(\mathbf{k}, \omega) = \int \frac{d\tau}{2\pi}\,e^{i\omega \tau}\,\int d^3\mathbf{x}\,e^{-i\mathbf{k}\mathbf{x}}\,T_{ij}(\tau, \mathbf{x}) \ .
\ee 
According to ~\cite{weinberg}, the total gravity wave energy per element of solid angle emitted by the localized source is given by
\be
\label{weinb}
\frac{d E_{\mathrm{gw}}}{d\Omega} = 2 G\,\mathcal{O}_{ij , lm}(\mathbf{\hat{k}})\,\int dk\,k^2\,T_{ij}(\mathbf{k}, k)\,T_{lm}^*(\mathbf{k}, k) \ .
\ee

For future reference, we now outline how this formula can be derived in our formalism. First, we set $a(\tau) = 1$ and take the limit 
$\tau_f \rightarrow \infty$ in Eq.~(\ref{rhof}). Next, we develop the sine and cosine into exponentials, and express 
$T_{ij}^{\mathrm{TT}}(\tau, \mathbf{k})$ in terms of $T_{ij}(\mathbf{k}, \omega)$ with (\ref{fourier}) and (\ref{TijTT}).
The resulting bilinear combination of $T_{ij}(\mathbf{k}, \omega)$ involves a bilinear combination of the projection operators 
$\mathcal{O}_{ij , lm}(\mathbf{\hat{k}})$, which satisfy
\be
\label{OO}
\mathcal{O}_{ij , lm}(\mathbf{\hat{k}}) \, \mathcal{O}_{ij , rs}(\mathbf{\hat{k}}) \, = \, \mathcal{O}_{lm , rs}(\mathbf{\hat{k}}) \ .
\ee 
Simplifying Eq.~(\ref{rhof}) for the gravity wave energy density accordingly, we recover Eq.~(\ref{weinb}) for the total energy in 
gravity waves per element of solid angle in $\mathbf{k}$-space.

The scale factors in our formula arise because, with the expansion of the universe, the energy density (\ref{rhof}) dilutes as 
radiation, $\rho_{\mathrm{gw}} \propto a^{-4}$, and the factor of $a(\tau')$ in the integrals comes from the fact that $\tau$ 
is the conformal time (and $k$ the comoving wave-number). Once again, note that in (\ref{Xk}) we may follow the evolution of 
gravity wave emission with the time $\tau_f$.

\subsection{Ensemble Average}
\label{ensav}

So far we have considered the emission of gravity waves by inhomogeneous media, without using their stochastic character.
We now consider the gravity wave energy density (\ref{rhogwmtw}) obtained from the ensemble average over different realizations of the 
random scalar fields. Proceeding as in (\ref{rhohbar}) and decomposing $h_{ij}(\tau, \mathbf{x})$ into Fourier components, 
we take the ensemble average $\langle ... \rangle$ of the resulting bilinear combination of the gravity wave amplitudes 
\be
\label{<hh>}
\rho_{\mathrm{gw}} = \frac{1}{32 \pi G a^4}\;\int \frac{d^3\mathbf{k}}{(2\pi)^{3/2}}\,\int \frac{d^3\mathbf{k'}}{(2\pi)^{3/2}}\; 
\langle \bar{h}'_{ij}(\mathbf{k}, \tau)\,\bar{h}_{ij}^{' *}(\mathbf{k'}, \tau) \rangle \; e^{i (\mathbf{k} - \mathbf{k'}) . \mathbf{x}} \ .
\ee
For $h_{ij}(\tau, \mathbf{k})$, we take again the free waves (\ref{hfree})-(\ref{AB}) after the emission process is completed. 
This involves the calculation of
\be
\label{<AA>}
\langle A_{ij}(\mathbf{k})\,A^*_{ij}(\mathbf{k'}) \rangle = \frac{(16\pi G)^2}{k\,k'}\,\int_{\tau_i}^{\tau_f} d\tau' \int_{\tau_i}^{\tau_f} d\tau''\,
\cos\left[k(\tau_f - \tau')\right]\,\cos\left[k'(\tau_f - \tau'')\right]\,a(\tau')\,a(\tau'')\;
\langle T_{ij}^{\mathrm{TT}}(\tau', \mathbf{k}) \, T_{ij}^{\mathrm{TT}*}(\tau'', \mathbf{k'}) \rangle \ .
\ee
We are thus led to consider the unequal time correlators of the transverse-traceless part of the energy momentum tensor (\ref{TijTT})
\ba
\label{<TT>}
&&
\hspace*{-1cm}
\langle T_{ij}^{\mathrm{TT}}(\tau', \mathbf{k}) \, T_{ij}^{\mathrm{TT}*}(\tau'', \mathbf{k'}) \rangle = 
\nonumber\\
&&\mathcal{O}_{ij , lm}(\mathbf{\hat{k}})\,\mathcal{O}_{ij , rs}(\mathbf{\hat{k'}})\,
\int \frac{d^3\mathbf{p}}{(2\pi)^{3/2}}\,\int \frac{d^3\mathbf{p'}}{(2\pi)^{3/2}}\,
p_l\,p_m\,p'_r\,p'_s \; \langle \phi_a(\mathbf{p}, \tau')\,\phi_a(\mathbf{k} - \mathbf{p}, \tau')\,\phi^*_b(\mathbf{p'}, \tau'')\,\phi^*_b(\mathbf{k'} - 
\mathbf{p'}, \tau'') \rangle \ .
\ea

For an arbitrary random medium, this depends on the 4-point unequal time correlators of the scalar fields. 
We can proceed further by assuming that the $\phi_a(\mathbf{p})$ are described by a statistically homogeneous random Gaussian field.
This description is often relevant in the context of preheating. The preheating process is conveniently divided
into four different stages: linear preheating, non-linear preheating with significance of backreaction, Kolmogorov turbulence
of scalar waves, and thermalization. Numerical investigations \cite{felder,bubbles,bubbles2} show that the scalar fields are Gaussian 
during the linear stage of preheating and non-Gaussian at the transient, ``bubbly'' stage and at the early turbulent stage,
while gaussianity is recovered at the later stage of turbulence and at the thermal stage. The following analytical analysis can be 
implemented for fields with Gaussian statistics. The 4-point functions may then be developed in terms of 2-point functions 
(cf. the Wick theorem) as
\ba
\label{<pppp>} 
\hspace*{-1cm}
\langle \phi_a(\mathbf{p}, \tau')\,\phi_a(\mathbf{k} - \mathbf{p}, \tau')\,\phi^*_b(\mathbf{p'}, \tau'')\,\phi^*_b(\mathbf{k'} - \mathbf{p'}, \tau'') \rangle 
&=& \langle \phi_a(\mathbf{p}, \tau')\,\phi^*_a(\mathbf{p} - \mathbf{k}, \tau') \rangle \; 
\langle \phi_b(- \mathbf{p'}, \tau'')\,\phi^*_b(\mathbf{k'} - \mathbf{p'}, \tau'')\rangle 
\nonumber \\
+ \, \langle \phi_a(\mathbf{p}, \tau')\,\phi^*_b(\mathbf{p'}, \tau'') \rangle \; 
\langle \phi_a(\mathbf{k} - \mathbf{p}, \tau')\,\phi^*_b(\mathbf{k'} - \mathbf{p'}, \tau'') \rangle 
&+& \langle \phi_a(\mathbf{p}, \tau')\,\phi^*_b(\mathbf{k'} - \mathbf{p'}, \tau'') \rangle \; 
\langle \phi_a(\mathbf{k} - \mathbf{p}, \tau')\,\phi^*_b(\mathbf{p'}, \tau'') \rangle
\ea
where we have used $\phi^*_a (\mathbf{p}) = \phi_a (-\mathbf{p})$ for real scalar fields. The unequal time correlators of the scalar fields are expressed as
\be
\label{<pp>}
\langle \phi_a(\mathbf{p}, \tau')\,\phi^*_b(\mathbf{p'}, \tau'') \rangle = F_{a b}(p, \tau', \tau'')\;\delta(\mathbf{p} - \mathbf{p'})
\ee
where $F_{ab}$ depends only on $p = |\mathbf{p}|$, by statistical homogeneity and isotropy of the scalar fields. 
The first term in the RHS of (\ref{<pppp>}) is proportional to $\delta(\mathbf{k})\,\delta(\mathbf{k'})$ and will not contribute 
(to the connected part of the 4-point function). Using (\ref{<pppp>}) with (\ref{<pp>}) in (\ref{<TT>}), we then get
\ba
\label{TTbis}
\hspace*{-1cm}
&& 
\langle T_{ij}^{\mathrm{TT}}(\tau', \mathbf{k}) \, T_{ij}^{\mathrm{TT}*}(\tau'', \mathbf{k'}) \rangle = 
\nonumber \\ 
\hspace*{-1cm}
&&
\delta(\mathbf{k} - \mathbf{k'})\,
\mathcal{O}_{ij , lm}(\mathbf{\hat{k}}) \, \mathcal{O}_{ij , rs}(\mathbf{\hat{k}}) \int \frac{d^3\mathbf{p}}{(2\pi)^3}
\left[p_l\,p_m\,p_r\,p_s + p_l\,p_m\,(k_r - p_r)\,(k_s - p_s)\right] F_{a b}(p, \tau', \tau'')\,F_{a b}(|\mathbf{k} - \mathbf{p}|, \tau', \tau'') 
\ea
where we have performed the integral over $\mathbf{p'}$ and enforced $\mathbf{k} = \mathbf{k'}$. 
In this expression, the terms proportional to $k_r$ and $k_s$ vanish when contracted with $\mathcal{O}_{ij , rs}(\mathbf{\hat{k}})$.
Using Eq.~(\ref{OO}), we then have
\be
\mathcal{O}_{lm , rs}(\mathbf{\hat{k})}\,p_l\,p_m\,p_r\,p_s = 
\frac{1}{2}\,\left[\mathbf{p}^2 - (\mathbf{\hat{k}} . \mathbf{p})^2\right]^2 = \frac{p^4}{2}\,\sin^4(\mathbf{\hat{k}}, \mathbf{\hat{p}})
\ee
where $(\mathbf{\hat{k}}, \mathbf{\hat{p}})$ denotes the angle between $\mathbf{k}$ and $\mathbf{p}$. Simplifying (\ref{TTbis}) accordingly, 
and inserting the 
result into (\ref{<AA>}), we get
\ba
\label{AAbis}
\langle A_{ij}(\mathbf{k})\,A^*_{ij}(\mathbf{k'}) \rangle &=&  \delta(\mathbf{k} - \mathbf{k'})\,\frac{(16\pi G)^2}{k^2}\,
\int_{\tau_i}^{\tau_f} d\tau' \int_{\tau_i}^{\tau_f} d\tau''\,\cos\left[k(\tau_f - \tau')\right]\,\cos\left[k(\tau_f - \tau'')\right]\,a(\tau')\,a(\tau'')
\nonumber \\ && 
\int \frac{d^3\mathbf{p}}{(2\pi)^3}\,p^4\,\sin^4(\mathbf{\hat{k}}, \mathbf{\hat{p}})\,
F_{a b}(p, \tau', \tau'')\,F_{a b}(|\mathbf{k} - \mathbf{p}|, \tau', \tau'')
\ea

Averaging over a complete period of oscillation of the free waves as in (\ref{timeav}), we then have
\be
\label{hhbis}
\langle \langle \bar{h}^{'}_{ij}(\mathbf{k}, \tau)\,\bar{h}_{ij}^{' *}(\mathbf{k'}, \tau) \rangle \rangle_{T}= 
\frac{k^2}{2}\,\left(\langle A_{ij}(\mathbf{k})\,A^*_{ij}(\mathbf{k'}) \rangle 
+ \langle B_{ij}(\mathbf{k})\,B^*_{ij}(\mathbf{k'}) \rangle \right)
\ee
where $\langle B_{ij}(\mathbf{k})\,B^*_{ij}(\mathbf{k'}) \rangle$ is given by (\ref{AAbis}) with the two cosines replaced by sines. 
These combine in (\ref{hhbis}) to give $\cos\left[k(\tau' - \tau'')\right]$. Finally, collecting all the factors in (\ref{<hh>}), 
and defining the gravity wave energy density spectrum as in (\ref{defSk}), we find
\be
\label{avSk}
S_k(\tau_f) = \frac{2}{\pi}\,G\,k^3 \int \frac{d^3\mathbf{p}}{(2\pi)^3}\,p^4\,\sin^4(\mathbf{\hat{k}}, \mathbf{\hat{p}})\,\int_{\tau_i}^{\tau_f} d\tau' 
\int_{\tau_i}^{\tau_f} d\tau'' \cos\left[k(\tau' - \tau'')\right]\,a(\tau')\,a(\tau'')\,F_{a b}(p, \tau', \tau'')\,
F_{a b}(|\mathbf{k} - \mathbf{p}|, \tau', \tau'')
\ee
where we have performed the integral over $\mathbf{k'}$ in (\ref{<hh>}), as well as the one over the direction $\mathbf{\hat{k}}$. 
This last step comes from the fact  that the result of the integral over $d^3\mathbf{p}$ in (\ref{avSk}) is
independent of the direction $\mathbf{\hat{k}}$, which is to say the gravity wave spectrum is isotropic. 

Eq.~(\ref{avSk}) is the main result of this section. We will use it in Section \ref{Analytic} and in sub-section \ref{analytics} to 
perform analytical calculations of the gravity wave energy density spectrum.

\subsection{Spectrum Today}
\label{speto}

Eventually we will be interested in the abundance of gravity wave energy density today
\be
\label{rhogwtoday}
h^2\,\left(\frac{\rho_{\mathrm{gw}}}{\rho_c}\right)_0 = \int \frac{df}{f}\,h^2\,\Omega_{\mathrm{gw}}(f) \ ,
\ee
and its spectrum per logarithmic frequency interval
\be
\label{spectoday} 
h^2\,\Omega_{\mathrm{gw}}(f) = \left(\frac{h^2}{\rho_c} \, \frac{d\rho_{\mathrm{gw}}}{d\ln f}\right)_0 \ ,
\ee
where $f$ is the frequency and $\rho_c = 3 H_0^2 / (8 \pi G)$ 
is the critical energy density today. To convert the gravity wave spectrum considered in the previous 
sections into physical variables today, we need to consider the evolution of the scale factor from preheating up to now, which 
depends on the evolution of the equation of state. In general, when the inflaton potential is quadratic around its 
minimum, the equation of state jumps from $w = 0$ to an intermediate value close to $w = 1/3$ during preheating \cite{eos}, \cite{tri}. 
We assume that the universe then evolves continuously towards radiation domination 
(e.g. without an intermediate matter dominated stage\footnote{An intermediate stage of matter domination could be due, for instance, 
to a massive relic dominating the expansion before it decays. If such a stage starts when the scale factor is $a(t) = a_1$, before the 
universe becomes radiation dominated at $a(t) = a_2$, then compared to the relations given below, the present-day gravity wave frequency 
$f$ would be smaller by a factor of $(a_1/a_2)^{1/4}$, and the amplitude of the spectrum $\Omega_{\mathrm{gw}}\,h^2$ would be smaller by 
a factor of $(a_1 / a_2)$.}). 

Let us denote by $t_i$ the end of inflation, by $t_j$ a moment after the jump of the equation of state (to be specified below), 
by $t_* > t_j$ the moment when thermal equilibrium is established, and by $t_0$ the present time. From $t_*$ to $t_0$ we assume 
the expansion of the universe obeys entropy conservation, and from $t_j$ to $t_*$ we assume that it evolves with a mean equation 
of state $w$ (which is thus close to $1/3$). The scale factor today compared to the one at the end of inflation may then be expressed as
\be
\label{ai/a0}
\frac{a_i}{a_0} = \frac{a_i}{a_j\,\rho_j^{1/4}}\,\left(\frac{a_j}{a_*}\right)^{1 - \frac{3}{4}(1+w)}
\,\left(\frac{g_*}{g_0}\right)^{-1/12}\,\rho_{\mathrm{rad} 0}^{1/4} \ ,
\ee   
where $\rho_j$ is the total energy density at $t=t_j$,  $\rho_{\mathrm{rad}0}$ is the energy density of radiation today, and $g$ is the number 
of effectively massless degress of freedom (neglecting the difference between $g$ and $g_S$), see \cite{kolbturner}. 
From Eq.~(\ref{ai/a0}), we deduce the physical wave-number today, $k_0 = k / a_0$, in terms of the the comoving wave-number $k$ used in the 
previous sections. The corresponding physical frequency today is given by
\be
f = \frac{k_0}{2\pi} = \frac{k}{a_j\,\rho_j^{1/4}}\,\left(\frac{a_j}{a_*}\right)^{1 - \frac{3}{4}(1+w)}\,4 \times 10^{10}\,\mathrm{Hz} \ .
\ee 
The energy density spectrum today is given by (\ref{defSk}) with $a = a_0$. Using (\ref{ai/a0}), this gives
\be
\Omega_{\mathrm{gw}}\,h^2 = \frac{h^2}{\rho_c} \, \frac{d\rho_{\mathrm{gw}}}{d\ln k} = 
\frac{S_k(\tau_f)}{a_j^4 \rho_j}\,\left(\frac{a_j}{a_*}\right)^{1-3 w}\,\left(\frac{g_*}{g_0}\right)^{-1/3}\,\Omega_{\mathrm{rad}}\,h^2 \ ,
\ee
where $\Omega_{\mathrm{rad}}\,h^2 = h^2\,\rho_{\mathrm{rad} 0} / \rho_c = 4.3 \times 10^{-5}$ is the abundance of radiation today, and we will 
take $g_* / g_0 = 100$. 

In the above relations, $a_j$ and $\rho_j$ may be derived directly form the simulations for a precise calculation. Everything is then known, 
except for the factor $a_j / a_*$, which depends on the unknown reheating temperature. Note that this factor reduces to one for $w = 1/3$. 
Therefore, by chosing $t_j$ such that $w_j$ is already close to $1/3$, this factor gives only a small contribution (which may be bounded by 
imposing some constraints on the reheat temperature $T_*$). 

In the case of a $\lambda\,\phi^4$ model of inflation, the average equation of state reaches $w = 1/3$ already during the simulation, in fact 
soon after the end of inflation, so the factor in $a_j / a_*$ is just absent in this case. The above relations then reduce to 
\be
\label{specphi4}
\mbox{For } \lambda\,\phi^4 \mbox{ model:} \;\;\;\;\; f = \frac{k}{a_j\,\rho_j^{1/4}}\;4 \times 10^{10}\,\mathrm{Hz} 
\;\;\;\;\; \mbox{ and } \;\;\;\;\; \Omega_{\mathrm{gw}}\,h^2 = 9.3 \times 10^{-6}\;\frac{S_k(\tau_f)}{a_j^4 \rho_j} \ ,
\ee
where $t_j$ will be chosen at the end of the simulations. Since in this model the average equation of state is the one of radiation very quickly 
after the end of inflation, we have $a_j^4\,\rho_j \simeq a_i^4\,\rho_i$, so that, normalizing the scale factor to one at the end of inflation, 
$a_j^4 \rho_j$ is essentially given by the energy density at that time, 
$a_j^4 \rho_j \simeq \lambda \, \phi_i^4 / 4 \simeq 3.4 \times 10^{-17}\,\Mp^4$ (for $\lambda = 10^{-14}$).

\section{Analytical Calculations of GW Emission}
\label{Analytic}

In this section we outline analytical calculations of the gravity wave spectrum emitted from preheating, and we illustrate the general 
formalism developed in Sections \ref{gwprod} and \ref{energyspectrum}. We consider four different stages of reheating: the linear preheating 
stage, the non-linear ``bubbly'' stage, scalar wave turbulence, and the stage of thermal equilibrium.

\subsection{Linear Preheating Stage}
\label{first}

Consider the first, linear stage of preheating when only one field, call it $\chi$, has been amplified.  
The quantum  field $\chi$ is described by the field operator 
\begin{equation}\label{osc}
\hat \chi(\tau, \mathbf{x}) = \int \frac{d^3\mathbf{k}}{(2\pi)^{3/2}}\,\left(\hat a_{\mathbf{k}}\,\chi_k(\tau)\,e^{ i{{\bf k}}{{\bf x}}}
+ \hat a^{+}_{\mathbf{k}}\,\chi^*_k(\tau)\,e^{ -i{{\bf k}}{{\bf x}}}\right) 
\end{equation}
where $\hat{a}^+_{\mathbf{k}}$ and $\hat{a}_{\mathbf{k}}$ are creation and annihilation operators with usual commutation relations. 
The field  $\hat \chi$ obeys Gaussian statistics, so that the formalism of Section \ref{ensav} should be applicable here.
In Eq.~(\ref{<pp>}), instead of the Fourier amplitude ${\chi}(\mathbf{p}, \tau)$, we have to use the operator\footnote{Note 
that the modes $\chi_p$ in (\ref{aa+}) have dimensions $\mathrm{mass}^{-1/2}$, while $\hat{\chi}(\mathbf{p}, \tau)$ has dimension 
$\mathrm{mass}^{-2}$.}
\be
\label{aa+}
{\chi}(\mathbf{p}, \tau)  \, \to \, \hat{\chi}(\mathbf{p}, \tau) \, \equiv \, 
\chi_p(\tau)\,\hat{a}_{\mathbf{p}} + \chi^*_p(\tau)\,\hat{a}^+_{-\mathbf{p}} \ . 
\ee
The field correlator is 
\be
\langle 0 | \hat{\chi}(\mathbf{p}, \tau')\,\hat{\chi}^+(\mathbf{p'}, \tau'') | 0 \rangle = \chi_p(\tau')\,\chi_p^*(\tau'')\,\delta(\mathbf{p} - \mathbf{p'})
\ee
which gives $F_{\chi \chi}(p, \tau', \tau'') = \chi_p(\tau')\,\chi_p^*(\tau'')$ in (\ref{<pp>}). Inserting this into Eq.~(\ref{avSk}), 
and expanding the cosine, the integrals over $\tau'$ and $\tau''$ are then complex conjugates of each other 
\ba
\label{avSkpre}
S_k(\tau_f) &=& \frac{2}{\pi}\,G\,k^3 \int \frac{d^3\mathbf{p}}{(2\pi)^3}\;p^4\,\sin^4(\mathbf{\hat{k}}, \mathbf{\hat{p}})
\nonumber \\
&& \left\{ \left| \int_{\tau_i}^{\tau_f} d\tau \, \cos\left(k\,\tau \right) \, a(\tau) \, \chi_p(\tau) \, \chi_{|\mathbf{k} - \mathbf{p}|}(\tau) \right|^2 
+ \left| \int_{\tau_i}^{\tau_f} d\tau \, \sin\left(k\,\tau \right) \, a(\tau) \, \chi_p(\tau) \, \chi_{|\mathbf{k} - \mathbf{p}|}(\tau) \right|^2 \right\} \ .
\ea

This equation allows us to perform analytical calculations of the gravity wave spectrum emitted in different models during the linear stage of 
preheating, where the time evolution of the eigenmodes $\chi_p(\tau)$ may be derived analytically. For example, in the case of broad parametric 
resonance with a quartic interaction $g^2\phi^2\chi^2$, the dispersion relation for the $\chi_p$ waves is 
\be
\label{disp1}
\omega_k=(k/a)^2+g^2\phi^2(t) \ ,
\ee
where $\phi(t)$ denotes the background inflaton oscillations at the end of chaotic inflation. For $\phi$ away from its zeros, 
the frequency $\omega_k$ is changing adiabatically with time, and the $\chi_p(\tau)$ are described by the WKB solution. According 
to the theorem of the sub-section \ref{nogw}, no gravity waves are emitted during these stages. However, when $\phi(t)$ crosses zero, 
$\omega_k$ changes non-adiabatically and gravity waves are emitted. We will calculate in detail the gravity wave spectrum emitted 
from broad parametric resonance below in sub-section \ref{analytics} for the model (\ref{lambdaphi4}). 
As we will see, our numerical results are reproduced very accurately from the analytical theory.

\subsection{Non-linear Bubbly Stage}
\label{second}

At the end of preheating, there is a violent, highly non-linear and non-perturbative stage, where the inhomogeneous fields have 
very large occupation numbers and strongly interact. The fields then become strongly non-Gaussian and analytical forms 
for their amplitudes are not available. Therefore, the analytical method of the previous section is not very useful for this stage. 
Yet, we expect this stage to give a dominant contribution to the gravity wave emission. It will thus be very useful to have an analytical 
estimate of the effects. 

For preheating after chaotic inflation, visualization of the field dynamics in configuration space \cite{bubbles}
reveals that, during the linear stage of preheating, scalar fields interacting with the inflaton are produced as standing random 
gaussian fields with exponentially increasing amplitude, then non-linear rescattering generates standing random non-gaussian 
inflaton inhomogeneities with very fast growing amplitude. The peaks of the inflaton inhomogeneities coincide with the peaks 
of the scalar fields produced by parametric resonance. When the inflaton peaks reach their maxima, they stop growing and begin 
to expand. The subsequent dynamics is characterized by the expansion and superposition of the scalar waves originated from the peaks. 
Multiple wave superposition results in phase mixing and sets up the turbulent waves dynamics. Thus, the short intermediate stage is 
characterized by the formation, expansion and collision of bubble-like fields inhomogeneities associated with the peaks of the
Gaussian field. This process is qualitatively similar to the bubble-like inflaton fragmentation occuring in tachyonic preheating 
after hybrid inflation \cite{tach1}, \cite{tach2}, \cite{tach3}.

The typical (physical) size $R_*$ of the bubble-like fields inhomogeneities when they fragment depends on the characteristic 
(comoving) momentum $k_*$ amplified by preheating, $R_* \sim a / k_*$. It was conjectured in \cite{bubbles} that this 
fragmentation gives the dominant contribution to gravity wave emission, with a total energy density in gravity waves of order
\be
\label{bubblescoll}
\left(\frac{\rho_{\mathrm{gw}}}{\rho_{\mathrm{tot}}}\right)_p \simeq  \, \alpha\,\left(R_*\,H\right)_p^2
\ee
at the time of production $t_p$, where $H$ is the hubble radius at that time (i.e. during the ``bubbly'' stage). 
We will estimate the coefficient  $\alpha$ from numerical simulations in sub-section \ref{numerics}. 
We will see that, for the model (\ref{lambdaphi4}), Eq.~(\ref{bubblescoll}) reproduces well the numerical calculations 
of the gravity wave emission, for $\alpha \sim 0.15$. Eq.~(\ref{bubblescoll}) seems rather general. It already appeared in the 
context of gravity wave emission from the collision of bubbles formed from first order phase transitions (in the thin-wall 
approximation)~\cite{Turner:1990rc}, \cite{Kosowsky:1991ua}. For comparison, in the case of the collision of two (relativistic) 
vacuum bubbles, Ref.~\cite{Kosowsky:1991ua} finds $\alpha = 1.3 \times 10^{-3}$.

\subsection{Scalar Wave Turbulence}
\label{third}

After preheating, the scalar fields enter a stage of turbulent interactions. The well-developed turbulent stage corresponds 
to media of the scalar waves, weakly interacting and having random phases. They have dispersion relations similar to (\ref{dispersion}) 
and are described by eigenfunctions varying as $e^{-i \omega_k t}$ with time. Thus, according to the no-go theorem of 
sub-section \ref{nogw}, no gravity wave emission is expected from the scalar waves' Kolmogorov turbulence. 
There are, however, some subtleties. In fact, numerical simulations of the scalar field dynamics show that developed
Kolmogorov turbulence is not established immediately after preheating ends. One of the indicators of this is that the 
fields stay non-Gaussian for a while after the end of preheating \cite{bubbles,bubbles2}. Also, for models of chaotic 
inflation, the residual inflaton condensate $\phi$ is still
significant after the end of preheating \cite{MT,eos}. 
Therefore, the wave-like dispersion relation (\ref{dispersion}) is not established immediately at this stage.
We shall expect some residual gravity wave production after the ``bubbly stage'' at the end of preheating. This is hard to 
treat analytically but can be seen in the numerical calculations of Section~\ref{lamphi4}.

\subsection{Thermal Bath}
\label{fourth}

Consider a scalar field $\varphi$ in thermal equilibrium with a thermal bath at a temperature $T$. 
As long as the temperature is higher than the Hubble parameter $H$, the expansion of the universe is not relevant.
The scalar field is described by the field operator decomposed into oscillators as in (\ref{osc}), which again obey Gaussian 
statistics. Here, instead of the Fourier transform of sub-section \ref{ensav}, we have to use the operator
\begin{equation}
\label{holon}
\varphi(\mathbf{k}, \tau) \to {e^{ -i\omega_{k}\tau} \over \sqrt{2\omega_{k}}} \, \hat a_{\mathbf{k}} 
+ {e^{ i\omega_{k}\tau} \over \sqrt{2\omega_{k}}} \, \hat a^{+}_{-\mathbf{k}}  \ .
\end{equation}
where $\omega_k^2 = k^2 + m^2$ and $m$ is the mass of $\varphi$. Then
we can directly use Eq.~(\ref{avSk}), where we have to 
calculate the unequal time field correlator (\ref{<pp>}). In order to do this, we  take an average for the thermal bath 
using the density matrix $Z^{-1}\, e^{-\hat H/T}$
\begin{equation}\label{bath}
\langle\varphi(\mathbf{k},\tau)\varphi^*(\mathbf{k}',\tau')\rangle = 
Z^{-1} Tr\left(e^{-\frac{\hat H}{T}}\hat \varphi(\mathbf{k},\tau)\hat \varphi^*(\mathbf{k}',\tau')\right) \ ,
\end{equation}
where the partition function $Z=\Pi_{\mathbf{k}}(1-e^{-\frac{\omega_{\mathbf{k}}}{T}})$ and the Hamiltonian for oscillators
$\hat H = \sum_{\mathbf{k}} \omega_{\mathbf{k}}  a^{+}_{\mathbf{k}} \hat a_{\mathbf{k}}$. Next, we substitute in (\ref{bath}) 
the operators in the form (\ref{holon}). Dropping the negligible contribution from the zero vacuum fluctuations, the result for 
the thermal, non-vanishing part is
\begin{equation}\label{details}
Z^{-1}\, Tr \left(e^{-\frac{\hat H}{T}} \hat a^{+}_{\mathbf{k}} \hat a_{\mathbf{k}'}  \right)=
\sum_{\mathbf{p}} \langle n_{\mathbf{p}}|e^{- \frac{\omega_{\mathbf{p}} n_{\mathbf{p}}}{T}} 
\hat a^{+}_{\mathbf{k}} \hat a_{\mathbf{k}'}|n_{\mathbf{p}}\rangle= \frac{\delta(\mathbf{k}-\mathbf{k}')}{e^{\frac{\omega_{\mathbf{k}}}{T}}-1} \ .
\end{equation}
Then, for a scalar field in equilibrium with a thermal bath at temperature $T$, the unequal time field correlator is
\begin{eqnarray}\label{eq:III-14}
\langle \varphi(\mathbf{p},\tau)\varphi^*(\mathbf{p}',\tau')\rangle =
\frac{\cos\left[\omega_p\,(\tau-\tau')\right]}{\omega_p\,\left(\hbox{e}^{\omega_p/T}-1\right)} \,
\delta(\mathbf{p}-\mathbf{p}') \ ,
\end{eqnarray}
so that 
\begin{equation}\label{eq:III-15}
F_{\varphi \varphi}(p, \tau, \tau') = \frac{\cos\left[\omega_p\,(\tau-\tau')\right]}{\omega_p\,\left(\hbox{e}^{\omega_p/T}-1\right)}
\end{equation}
in Eq.~(\ref{<pp>}).

The field correlator is a harmonic function of time. Therefore after inserting it in Eq.~(\ref{avSk}), we again
reproduce the conditions for the no-go theorem of sub-section \ref{nogw}. No gravity waves are emitted from the 
thermal bath of scalar fields. However, the bath of gauge fields and photons emits gravity waves \cite{ford}.

\section{Numerical Methods for GW Calculation}
\label{Numerical}

In this Section we describe our numerical algorithm, which is based on the formalism of
Section~\ref{energyspectrum}. Then we compare our method with previous numerical calculations 
of gravity waves produced from preheating.

\subsection{Numerical Algorithm}
\label{numethod}

\begin{figure}[htb]
\begin{minipage}[t]{8cm}
\centering \leavevmode \epsfxsize=5cm
\epsfbox{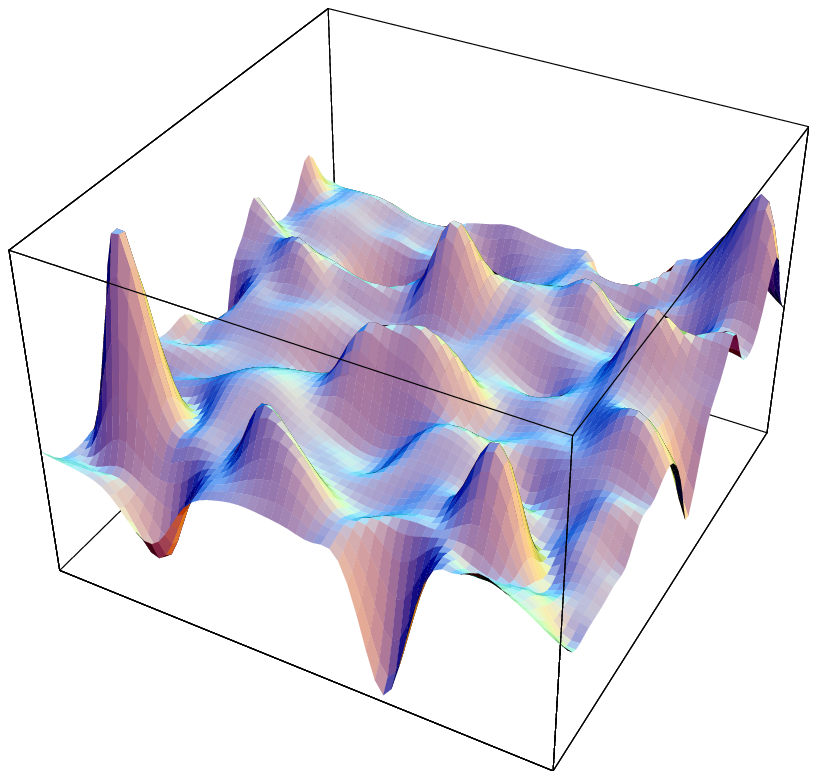}
\caption{Two-dimensional slice through a three-dimensional realization of the 
scalar field $\chi$ from a numerical simulation of preheating in the $\lambda \phi^4 +g^2\phi^2 \chi^2$ model
of the next section. The horizontal axes correspond to spatial coordinates and the vertical axis corresponds 
to the field's values.}
\label{fig:field}
\end{minipage}
\hspace*{1.5cm}
\begin{minipage}[t]{8cm}
\centering \leavevmode \epsfxsize=8cm
\epsfbox{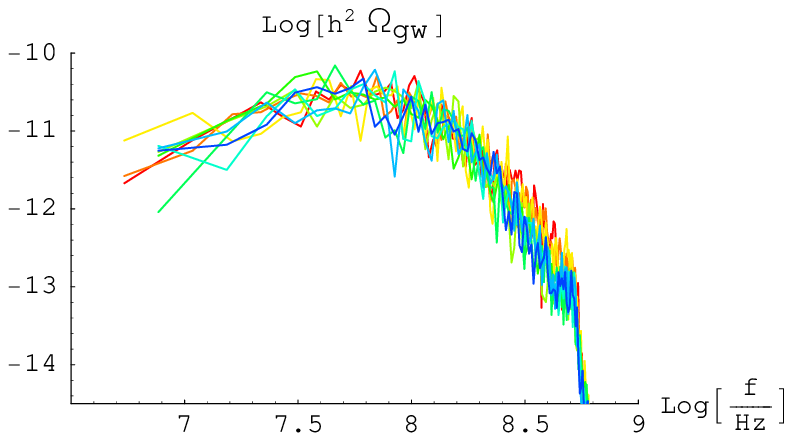}\\
\caption{Spectrum of energy density in gravity waves calculated along nine different directions 
in $\mathbf{k}$-space. The parameters for this run were the same as for Fig.~\ref{omegafq120}, 
see sub-section \ref{numerics} for details.}
\label{fig:dir}
\end{minipage}
\end{figure}

Our method is based on numerical computation of the quantity
$S_k(\tau_f)$ defined in (\ref{Xk}). We use the program LATTICEEASY
\cite{LE} to calculate the evolution of the scalar fields and the evolution of the 
scale factor. A typical realization of the scalar field profile in the most interesting
``bubbly'' stage (during preheating) is shown in the
Figure~\ref{fig:field}. The calculation of the scalar field evolution is performed 
in configuration space and is described in the LATTICEEASY documentation. We added a 
function to periodically calculate the integrands in equation (\ref{Xk}). 

To find the integrand at each integration time step we calculated the
transverse, traceless components of the scalar fields' energy momentum
tensor $T_{ij}^{\mathrm{TT}}(\tau', \mathbf{k})$, given by equation
(\ref{TijTT}). For each field we calculated six arrays with the mixed
derivatives $\left\{\partial_i \phi \, \partial_j \phi\right\}(\mathbf{x})$ 
in configuration space, and then Fourier Transformed them to obtain
the six arrays $\left\{\partial_i \phi \, \partial_j \phi\right\}(\mathbf{k})$.

We found the projection operator $\mathcal{O}_{ij , lm}$ analytically
in each direction we used, using Eqs.~(\ref{TTproj}) and
(\ref{Pij}). For example, from Eq.~(\ref{Pij}) we can see that in the
$k_x$ direction the only nonvanishing components of $P_{ij}$ are
$P_{22}=P_{33}=1$. From Eq.~(\ref{TTproj}) you can thus show that the
only nonvanishing components of $\mathcal{O}_{ij , lm}$ along this
axis are $\mathcal{O}_{22 , 22} = \mathcal{O}_{33 , 33} = (1/2)$,
$\mathcal{O}_{22 , 33} = \mathcal{O}_{33 , 22} = -(1/2)$,
$\mathcal{O}_{23 , 23} = \mathcal{O}_{32 , 32} = 1$. Finally, we can
plug this into Eq.~(\ref{Tscal}) to find the only nonvanishing
components of $T_{ij}^{\mathrm{TT}}(\tau', \mathbf{k})$ along the
$k_x$ axis:
\begin{eqnarray}
T_{22}^{\mathrm{TT}} = -T_{33}^{\mathrm{TT}} &=& {1 \over 2}
\left[\left\{\partial_2 \phi \, \partial_2 \phi\right\}(\mathbf{k}) -
  \left\{\partial_3 \phi \, \partial_3 \phi\right\}(\mathbf{k})\right]
\\
\nonumber T_{23}^{\mathrm{TT}} = T_{32}^{\mathrm{TT}} &=&
\left\{\partial_2 \phi \, \partial_3 \phi\right\}(\mathbf{k})
\end{eqnarray}
In a similar way one can derive expressions for the components of
$T_{ij}^{\mathrm{TT}}(\tau', \mathbf{k})$ along any direction in $k$
space.

We calculated the values of $S_k$ along nine different directions
in Fourier space. The first three were the three axes $k_x$, $k_y$,
and $k_z$ and the remaining six were directions at $45^\circ$ in
between two of the axes: $xy$, $xz$, $yz$, $(-x)y$, $(-x)z$, and
$(-y)z$. The results for the spectrum $\Omega_{gw} h^2$ calculated 
along these nine
different directions is shown in Figure~\ref{fig:dir}. This figure
shows that $S_k$ is nearly identical in different $\mathbf{k}$-directions 
(i.e. the gravity wave spectrum is isotropic), but it also shows that there 
is statistical noise in each spectrum plot. Therefore, we averaged over the six diagonal
directions, which significantly reduces this noise. Unless otherwise noted all spectra shown 
in this paper represent averages over the six diagonal directions in $k$ space. (We
did not include the axes because in a cubic lattice the $k$ values
along the axes are different from the ones along diagonals, making
it harder to average the results together.) In Figures
\ref{omegafq120} and \ref{check} an average over the three axial
directions is plotted as well as an average of the six diagonal
directions.

Finally, we used Eq.~(\ref{specphi4}) to calculate $f$ and
$\Omega_{gw} h^2$ today. We confirmed that the combination $a_j^4
\rho_j$ was constant from a time shortly after inflation to the end of
the simulation, as it should be for radiation domination, so we used
the values from the end of the simulation.

\subsection{Comparison with Other Methods}
\label{comparison}

In the literature, there are different interesting methods of numerical calculation of gravity waves emission 
from preheating. In this sub-section, we contrast our formalism to previous numerical methods. 

Gravitational waves from preheating after chaotic inflation were
investigated in \cite{pregw1}, \cite{pregw2}, on the basis of the Weinberg 
formula (\ref{weinb}) in Minkowski spacetime. The expansion of the universe was then taken into account by dividing the conformal 
time into steps $\Delta \tau$, calculating the gravity wave energy density resulting from each step, and summing up each contribution 
diluted by the value of the scale factor at the middle of the step. In addition to the crude account of the evolution of the scale 
factor, one may show from sub-section \ref{weinberg} that such a treatment effectively replaces the terms
\be
\label{correct}
\left| \int_{\tau_i}^{\tau_f} d\tau'\,\cos\left(k\,\tau'\right)\,a(\tau')\,T_{ij}^{\mathrm{TT}}(\tau', \mathbf{k}) \right|^2
\ee
in Eq.~(\ref{Xk}) by
\be
\label{wrong}
\sum_{\alpha} \left| \int_{\tau_{\alpha}}^{\tau_{\alpha} + \Delta \tau} d\tau'\,\cos\left(k\,\tau'\right)\,a(\tau')\,
T_{ij}^{\mathrm{TT}}(\tau', \mathbf{k}) \right|^2 \ ,
\ee
where $\alpha=1,2,3, ...$ counts the steps. Indeed, what is summed up in Refs.~\cite{pregw1}, \cite{pregw2} 
is the energy density from each partial step, instead of the energy density from the whole evolution of gravity waves with time. 
To compare this with our method, we can divide the time integral in (\ref{correct}) into similar steps. Obviously, 
the result is different from (\ref{wrong}), which  misses all the cross terms given by the products of integrals over 
different steps. In other words, the approach based on (\ref{wrong}) does not take into account the precise propagation 
history of the gravitational waves in the medium. Nevertheless, this may give a good approximation during preheating 
itself, when the scalar fields are exponentially amplified and their past values give negligible contributions. However, 
the past evolution should be taken into account at the end of preheating, when the amplitude of the scalar fields stabilize 
around their maximum value. As we will see in the next section, our numerical results are relatively similar to the ones 
obtained in the earlier paper \cite{pregw1} (where a smaller frequency range is considered), but they differ significantly from the results 
of \cite{pregw2}, for both the shape and the amplitude of the gravity wave 
spectrum (see sub-section \ref{numerics}).

Another method to calculate the production of gravitational waves from preheating was used recently in 
\cite{pregw4}, in the context of hybrid inflation models, without expansion of the universe. 
Ref.~\cite{pregw4} considers the traceless part $\tilde{h}_{ij}$ ($\tilde{h}_{ii} = 0$) of the spatial components of 
the linear metric perturbation in the harmonic (de Donder) gauge ($\partial^{\mu} h_{\mu\nu} = 0$), satisfying  
\be
\label{tildeh}
\tilde{h}_{ij}'' - \mathbf{\nabla}^2 \tilde{h}_{ij} = 16\pi G\;
\left[\partial_i \phi\,\partial_j \phi - \frac{\delta_{ij}}{3}\,\partial_k \phi\,\partial_k \phi\right]
\ee
and solve for this equation in configuration space. Gravity waves correspond to the two independent degrees of freedom 
of the transverse and traceless part of the spatial components of the metric perturbations, i.e. the tensor part, 
that we denote $h_{ij}$. However, in addition to the tensor part $h_{ij}$, the five independent degrees of freedom of the traceless 
$\tilde{h}_{ij}$ in harmonic gauge involve an extra scalar part (with one degree of freedom) and an extra (divergence-less) 
vector part (with two degrees of freedom). In configurations space, one has to solve for these extra components of the metric perturbation 
as well, because the transverse-traceless projection (\ref{TTproj}) is then non-local. Only the transverse-traceless part contributes 
to the energy density in gravity waves, and therefore it has to be extracted efficiently\footnote{In Weinberg's formalism 
in sub-section~\ref{weinberg}, which is also derived in the harmonic gauge, the separation of the gravity waves from the other 
modes is achieved by the wave zone approximation far away from the source and by imposing the dispersion relation 
$\omega_k=|\mathbf{k}|$ for the solution of the wave equation.}. Ref.~\cite{pregw4} proposes to use the formula 
\be
\label{dani}
\left( \rho_{\mathrm{gw}} \propto \right) \;\; \langle \dot{h}_{ij}(\mathbf{x})\,\dot{h}_{ij}(\mathbf{x}) \rangle = 
\frac{2}{5}\, \langle \dot{\tilde{h}}_{ij}(\mathbf{x})\,\dot{\tilde{h}}_{ij}(\mathbf{x}) \rangle
\ee
to relate the energy density in gravity waves to the spatial average of all the components of the traceless 
$\dot{\tilde{h}}_{ij}$. 

We can justify Eq.~(\ref{dani}) in the special case of gravity waves emitted far away from an isolated source, in the quadrupole 
approximation (i.e. for wavelengths large compared to the size of the localized source). Indeed, Eq.~(\ref{tildeh}) can be 
solved in Fourier space with the Green function similar to the solutions (\ref{h}), (\ref{h'}) (here in Minkowski spacetime)
\be
\label{hh}
\dot{\tilde{h}}_{ij}(t, \mathbf{k}) = \int_{t_i}^{t} dt\,\cos\left[k\,(t - t')\right]\,M_{ij}(t', \mathbf{k})  \ ,
\ee
where $M_{ij}(t, \mathbf{k})$ is the Fourier transform of the RHS of Eq.~(\ref{tildeh}). The transverse-traceless $h_{ij}$ 
is then obtained from the traceless $\tilde{h}_{ij}$ by the projection (\ref{TTproj}) in Fourier space, which satisfies 
(\ref{OO}). The spatial average of the bilinear combination of the tensor modes involved in the gravity wave energy density 
is then given by
\be
\label{dani2}
\langle \dot{h}_{ij}(t, \mathbf{x})\,\dot{h}_{ij}(t, \mathbf{x}) \rangle = \int \frac{d^3\mathbf{k}}{(2\pi)^{3/2}}\, 
\mathcal{O}_{lm , rs}(\mathbf{\hat{k}}) \, \dot{\tilde{h}}_{lm}(t, \mathbf{k}) \, \dot{\tilde{h}}_{rs}(t, \mathbf{k}) \ .
\ee
In the quadrupole approximation, the Fourier transform of the source, $M_{ij}(t, \mathbf{k})$, does not depend on $\mathbf{k}$, 
so that $\dot{\tilde{h}}_{ij}(t, \mathbf{k})$ in (\ref{hh}) does not depend on the direction $\mathbf{\hat{k}}$. In this case, 
the integrand in the RHS of Eq.~(\ref{dani2}) depends on the direction $\mathbf{\hat{k}}$ only through $\mathcal{O}_{lm , rs}(\mathbf{\hat{k}})$, 
giving a $2/5$ factor after integration over the solid angle in $\mathbf{k}$-space. This then leads to Eq.~(\ref{dani}). 
However, it is not clear how this equation is applicable outside the quadrupole approximation and for extended sources. 

Indeed we can give an example of scalar fields configurations relevant for preheating where the method of \cite{pregw4} 
is not applicable. As in sub-section~\ref{nogw}, consider media of superposed scalar field waves. 
There is a gravitational response to its inhomogeneities 
and dynamics in the form of scalar modes, but the tensor modes are not emitted in this case, as we derived in  
sub-section~\ref{nogw}. Indeed for this case, in (\ref{dani}) we have 
$\langle \dot{h}_{ij}(\mathbf{x})\,\dot{h}_{ij}(\mathbf{x}) \rangle \, = \, 0$, 
but in general (except in the quadruple approximation) 
$\langle \dot{\tilde{h}}_{ij}(\mathbf{x})\,\dot{\tilde{h}}_{ij}(\mathbf{x}) \rangle \, \neq \, 0$. This may be checked explicitely 
with the solution (\ref{hh}) by proceeding as we did in sections \ref{nogw} and \ref{energyspectrum}. The argument of sub-section 
\ref{nogw} does not apply for $\tilde{h}_{ij}$ (in other words, trilinear interactions between $\tilde{h}_{ij}$ and two scalar particles 
do not violate helicity conservation), because it involves a scalar degree of freedom. The method of \cite{pregw4} is thus not generic, 
since it would lead to the incorrect conclusion that there is a non-vanishing emission of gravity waves from the media of scalar waves.


\section{Application to Preheating in the $\lambda \, \phi^4$ Model}
\label{lamphi4}


In this section, we illustrate our results for a simple model of preheating after $\lambda\,\phi^4$ chaotic inflation. 
We first review some useful features of the scalar field dynamics. Then we present our numerical results for the 
spectra of energy density in gravitational waves. Finally, we give an analytical treatment of the gravity wave production 
during the linear stage of preheating, which allows us to check our numerical results in detail.

\subsection{The Model}
\label{model}

The model we consider corresponds to two scalar fields minimally coupled to gravity with the potential
\be
\label{lambdaphi4}
V = \frac{\lambda}{4}\,\phi^4 + \frac{g^2}{2}\,\phi^2\,\chi^2
\ee
where $\lambda = 10^{-14}$ (from normalisation of CMB anisotropies). The theory of preheating for this model was developed 
in \cite{greene}. Because of conformal invariance, the expansion of the universe may be scaled out of the equations of motion. 
It is convenient to redefine the fields and the time coordinate as
\be
\label{redef}
X = a\,\chi \;\;\;\; , \;\;\;\; \Phi = a\,\phi \;\;\;\; , \;\;\;\; x = \sqrt{\lambda}\,\phi_0\,\tau
\ee
where $\phi_0 \simeq 0.342\,\Mp$ is the initial amplitude of the inflaton condensate at the end of inflation, and the scale factor $a$ 
is normalised to $1$ at that time. During the linear stage of preheating (when backreaction is negligible), 
the temporal part of the modes $X_k(x)$ obey the oscillator equation\footnote{In fact, the actual eigenmode equation (\ref{modesequ}) involves a term 
$\frac{a''}{a}$ in the effective frequency. For the background inflaton oscillating around the minimum of the 
quartic potential, the average value of $\frac{a''}{a}$ is zero. However, immediately after inflation, this term gives 
some residual contribution to the effective frequency, which slightly affects the solutions for the eigenmodes and for 
the background oscillations.}
\be
\label{modesequ}
\frac{d^2 X_k}{dx^2} + \omega_k^2(x)\,X_k = 0 \;\;\;\; \mbox{ with } \;\;\;\; \omega_k^2(x) = K^2 + q\,\bar{f}^2(x) 
\ee
where $\bar{f}$ denotes the inflaton zero-mode rescaled by its initial amplitude, $\bar{f} = \bar{\Phi} / \phi_0$. 
At this stage, it is given by a Jacobi elliptic cosine, $\bar{f}(x) = \mathrm{cn}[x, 1/2]$. We have also defined
\be
\label{Kandq}
K = \frac{k}{\sqrt{\lambda}\,\phi_0} \;\;\;\; , \;\;\;\; q = \frac{g^2}{\lambda}
\ee
where $k$ denotes comoving momenta.

During preheating, the modes are amplified by parametric resonance as the inflaton $\bar{\Phi}$ oscillates around the minimum 
of the potential. The key parameter here is the resonance parameter $q$. Since the coupling $\lambda$ is very small, we may 
expect that $q >> 1$. The modes $X_k$ are then amplified in small intervals of time, when the inflaton crosses zero. Parametric 
resonance amplifies only infrared modes belonging to resonance bands. For large $q$, the typical momenta $k_*$ and widths $\Delta k$
of the resonance are bounded by~\cite{greene}
\be
\label{kstar}
k_* \, , \, \Delta k  \stackrel{<}{\sim}  q^{1/4}\,\sqrt{\lambda}\,\phi_0
\ee 
Note that the inflaton fluctuations $\Phi_k$ obey an equation similar to the one for $X_k$, but with $q = 3$. The resonance is 
not very efficient in this case, so that generally the $\chi$-modes are amplified first. However, as soon as the $\chi$-field 
reaches high occupation numbers, it excites inhomogeneous inflaton fluctuations through its interaction with $\phi$, again with 
very large amplitude. 

When the amplitude of the fields is sufficiently large, the dynamics becomes fully non-linear, but it may be investigated with the 
help of lattice simulations~\cite{tkachev}, \cite{Prokopec:1996rr}. This is the stage we will mainly consider in the next section 
when we present our numerical results. Although the details of preheating depend on the particular model considered, the non-linear 
evolution exhibits several generic features~\cite{felder}. At the end of preheating, there is a violent stage of rescattering, with 
a restructuring of the scalar field spectra~\cite{tkachev}, \cite{rescattering}. Around the same time, large bubble-like field 
inhomogeneities, amplified during preheating, collide to form smaller structures \cite{bubbles}. Finally, there is a much longer 
regime of turbulent interactions, during which the spectra slowly propagate towards thermal equilibrium.

\subsection{Numerical Results}
\label{numerics}

We computed the gravitational radiation emitted in the model (\ref{lambdaphi4}) for several values of the coupling constant $g^2$. 
The abundance of gravity wave energy density today and its spectrum per logarithmic frequency interval were calculated numerically 
from Eqs.~(\ref{specphi4}) and (\ref{Xk}), as outlined in section \ref{numethod}. For the model (\ref{lambdaphi4}), the universe is 
radiation dominated soon after the end of inflation, so that $a_j^4 \, \rho_j$ in Eq.~(\ref{specphi4}) is very close to the initial 
energy density at the end of inflation (where the scale factor is normalised to unity). The precise value was taken from the simulations: 
$a_j^4\,\rho_j \simeq 1.15\,\lambda\,\phi_0^4 / 4$, for all the cases we considered.  

\begin{figure}[htb]
\begin{minipage}[t]{8cm}
\centering \leavevmode \epsfxsize=8cm
\epsfbox{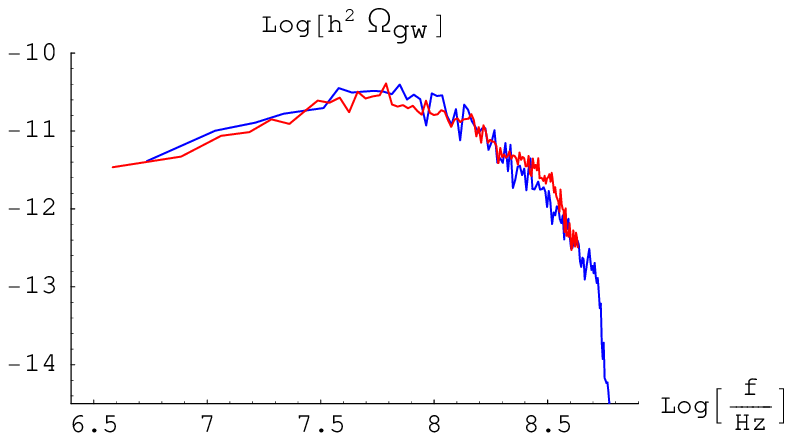}
\caption{Spectrum of gravity wave energy density in physical variables today (\ref{spectoday}), accumulated up to the time $x_f = 240$, 
for the model (\ref{lambdaphi4}) with $q = 120$. The 2 spectra were obtained from simulations with different box sizes, and averaged 
over different directions in $\mathbf{k}$-space.}
\label{omegafq120}
\end{minipage}
\hspace*{1.5cm}
\begin{minipage}[t]{8cm}
\centering \leavevmode \epsfxsize=8cm
\epsfbox{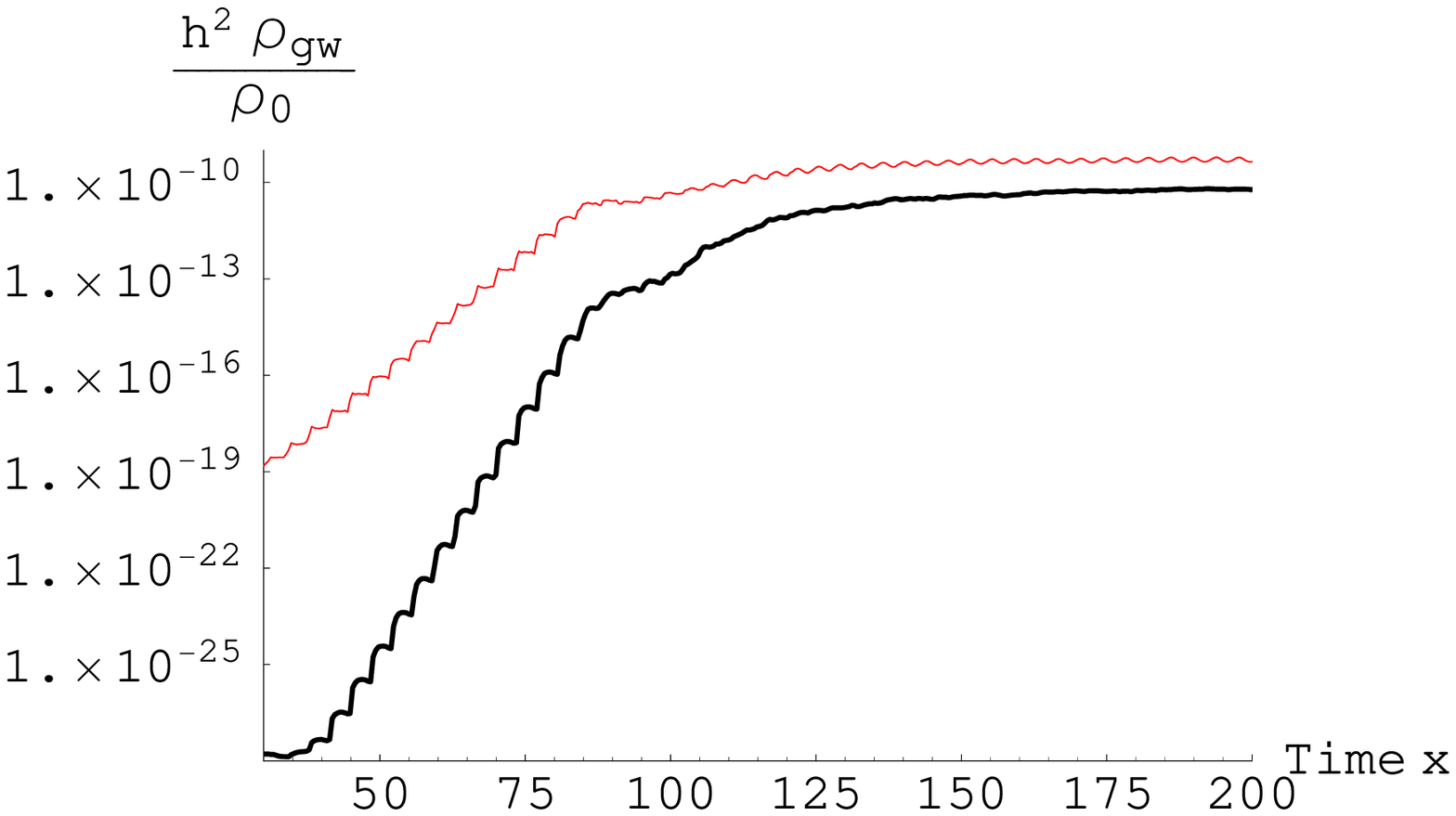}
\caption{The thick curve shows the total energy density in gravity waves (\ref{rhogwtoday}) accumulated up to the time $x_f$, 
as a function of $x_f$. The thin curve shows the evolution with time of the total particles 
number density, $n_{\mathrm{tot}} = n_{\chi} + n_{\phi}$, 
rescaled to fit on the same figure.}
\label{rhogwtime}
\end{minipage}
\end{figure}

\begin{figure}[htb]
\begin{minipage}[t]{8cm}
\centering \leavevmode \epsfxsize=8cm
\epsfbox{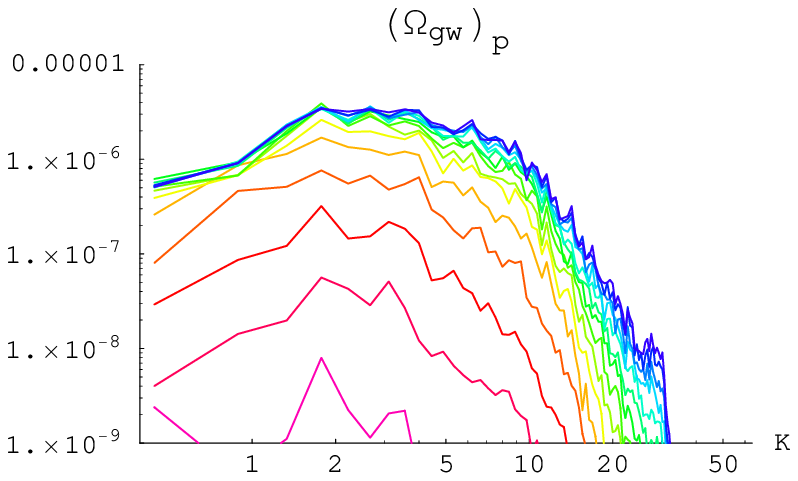}
\caption{Spectrum (\ref{specprod}) of the gravity wave energy density, accumulated up to different times $x_f$, as a function of the 
comoving momentum $k$ (in units of $\lambda \phi_0$). The spectra are shown from $x_f = 90$ to $x_f = 240$ with spacing $\Delta x_f = 10$.}
\label{omegatime}
\end{minipage}
\hspace*{1.5cm}
\begin{minipage}[t]{8cm}
\centering \leavevmode \epsfxsize=8cm
\epsfbox{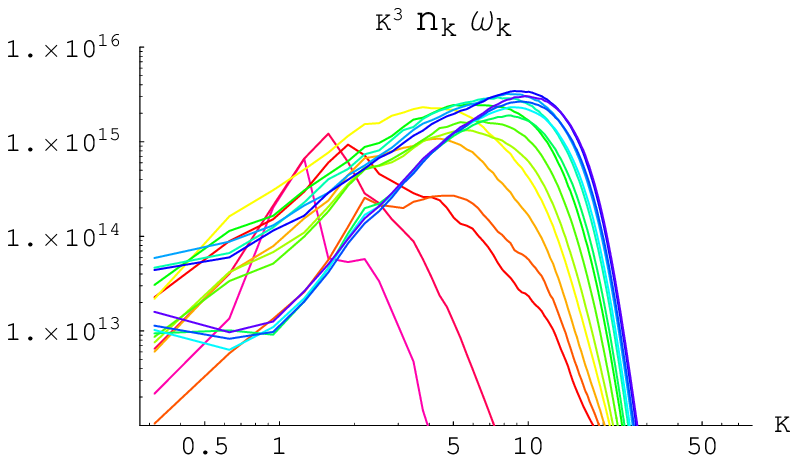}\\
\caption{Measure of the (unnormalised) total energy density in the two scalar fields per logarithmic momentum interval at different moments 
of time. The same times as in Fig.~(\ref{omegatime}) are shown, the spectra moving towards UV from $x = 90$ to $x = 240$ with spacing $\Delta x = 10$.}
\label{enpermode}
\end{minipage}
\end{figure}

We first present the results of simulations for $q = g^2/\lambda = 120$. Fig.~\ref{omegafq120} shows the gravity wave spectrum 
(\ref{spectoday}), accumulated up to the time $x_f = 240$ ($\tau_f = 240 / (\sqrt{\lambda}\,\phi_0)$ in (\ref{Xk})), i.e. 
well after the end of the preheating stage. Each of these spectra were
calculated by averaging over different directions in $\mathbf{k}$-space as described
in \ref{numethod}, but the directions used were different for the two
plots. The two plots were also obtained from simulations with different box sizes. 
Fig.~\ref{omegafq120} thus confirms that the results are isotropic and independent of
the box volume. For this model the peak amplitude was $h^2\,\Omega_{\mathrm{gw}}
\sim 3 \times 10^{-11}$ with a peak frequency 
$f \sim 7 \times 10^7\,\mathrm{Hz}$. The infrared tail of the spectrum at the end of the simulation varied approximately as 
$\Omega_{\mathrm{gw}} \propto f$. Of course, the spectrum is expected to decay much faster at lower frequencies corresponding 
to modes produced outside the Hubble radius. For this model, the Hubble rate dropped from $H \sim 2.5 \times 10^{-8}\,\Mp$ 
at the end of inflation to $H \sim 5 \times 10^{-11}\,\Mp$ at the beginning of preheating, and then slowly decreased up to 
$H \sim 10^{-12}\,\Mp$ at the end of the simulation. For $k/a = H$ at the time of production, this gives a frequency today: 
$f \sim 10^7\,\mathrm{Hz}$, $f \sim 5 \times 10^5\,\mathrm{Hz}$ and $f \sim 5 \times 10^4\,\mathrm{Hz}$ respectively.
In particular, a physical momentum of the order of the inverse Hubble radius when the main part of the spectrum is produced 
corresponds to a frequency today $f \sim 10^5\,\mathrm{Hz}$. All the modes shown in Fig.~(\ref{omegatime}) were thus well inside the 
Hubble radius at the time of production. This also means that any
gravitational waves produced from preheating in this model would be at
frequencies well above the range of LIGO/VIRGO (to say nothing of LISA or DECIGO/BBO).

Figure~\ref{omegafq120} can be directly compared with Figure~1 of \cite{pregw2}
which corresponds to the same parameters and the same frequency range. The spectrum 
obtained in Ref.~\cite{pregw2} has a peak amplitude of $h^2\,\Omega_{\mathrm{gw}} \sim 10^{-9}$  
at the frequency $f \sim 10^{8.3} Hz$. In our case, the peak of the spectrum occurs at a lower frequency, 
and its amplitude is smaller by factor of order $1 / 30$. The spectrum we obtain is broader, and in particular 
it decreases more slowly in the infrared. In the next section we will confirm our results analytically. 

We now consider in more detail the role of the different stages of the scalar field evolution on the production of gravity waves. 
Figure~\ref{rhogwtime} shows how the total energy density in gravity waves (\ref{rhogwtoday}) is accumulated with time. For comparison, 
we plot on the same figure the evolution with time of the total particle 
number density of $\chi$ and $\phi$. The first stage on this plot, up 
to $x_f \simeq 90$, corresponds to preheating by parametric resonance. During this stage, gravity waves are produced in very short 
intervals where the scalar fields are non-abiatically amplified. A detailed discussion of this regime will be given in the next sub-section. 
The resulting gravity wave energy density grows with twice the exponent of the scalar field number density, 
$\rho_{\mathrm{gw}} \propto n_{\mathrm{tot}}^2$. At the end of preheating, during the short, highly non-linear stage of ``rescattering'', the 
energy density in gravity waves continues to increase significantly up to $x_f \simeq 150$, when the number density becomes approximately 
constant. For different values of the $q$-parameter, this rescattering stage may be much shorter, and hardly distinguishable form the end 
of preheating. Finally, the gravity wave energy density still slightly increases during the early stage of
the subsequent turbulent regime, but at a much lower rate. As shown in Figure~\ref{rhogwtime}, the exponential rate 
of gravity wave production is maximal during parametric resonance, but the main part of the final energy density in 
gravity waves is produced during the ``bubbly'' stage, a short intermediate stage between parametric resonance and the onset of 
the turbulent regime.   

Let us now discuss how the gravity wave spectrum builds up with time. Here we will consider the spectrum of the energy density in 
gravity waves around the time of their production
\be
\label{specprod}
\left(\Omega_{\mathrm{gw}}\right)_p = \left(\frac{1}{\rho_{\mathrm{tot}}}\,\frac{d \rho_{\mathrm{gw}}}{d \ln k}\right)_p 
\simeq \frac{S_k(\tau_f)}{a_j^4\,\rho_j}
\ee
as a function of the comoving momentum $k$. Remember that $a_j^4\,\rho_j \simeq 1.15\,\lambda\,\phi_0^4/4$. 
The spectrum (\ref{specprod}), accumulated up to different times $\tau_f$, is shown in Fig.~(\ref{omegatime}). For a better view, we only 
plot the spectra starting from the end of preheating up to the end of the simulation, because the final spectrum is mainly determined by this 
stage (see Fig.~\ref{check} of the next sub-section for a plot of the spectrum during the preheating stage). For comparison, we show in 
Fig.~\ref{enpermode} the spectra of the quantity $k^3\,\omega_k^{\chi}\,n_k^{\chi} + k^3\,\omega_k^{\phi}\,n_k^{\phi}$ 
at the same moments of time, where $n_k$ and $\omega_k$ are the occupation number and the frequency of the scalar modes. This quantity 
gives a rough measure of the total energy density in the scalar field fluctuations per logarithmic momentum 
interval (provided $\omega_k$ is changing nearly adiabatically), in analogy with (\ref{specprod}) for the gravitational waves. During preheating, 
the scalar field spectra are strongly peaked around the resonant momenta inside the resonance bands. The gravity wave spectrum has a peak 
amplitude growing exponentially up to $(\Omega_{\mathrm{gw}})_p \sim 10^{-9}$ at $x_f \sim 90$. The subsequent evolution is shown in 
Figs.~\ref{omegatime}, \ref{enpermode}. The peak of the scalar field spectrum 
is initially in the infrared, around the resonant momentum of preheating, and then quickly moves towards the ultraviolet as a result of 
rescattering. The gravity wave spectrum increases significantly during
this stage, but in contrast to the scalar fields, the frequency of its peak 
stays approximately the same. The maximal gravity wave amplitude ($(\Omega_{\mathrm{gw}})_p \sim 5 \times 10^{-6}$) is reached around 
$x_f \sim 150$. Later on, both spectra slowly propagate towards higher frequencies, which slightly increases the total energy density 
in gravitational waves. On the other hand, the peak amplitude and the infrared part of the gravity wave spectrum stay unchanged after 
$x_f \sim 150$. (We checked this with simulations extending further in the turbulent
regime). Their final value may thus be efficiently determined by finite-time simulations.

\begin{figure}[hbt]
\begin{center}
\begin{tabular}{cc}
\includegraphics[width=7cm]{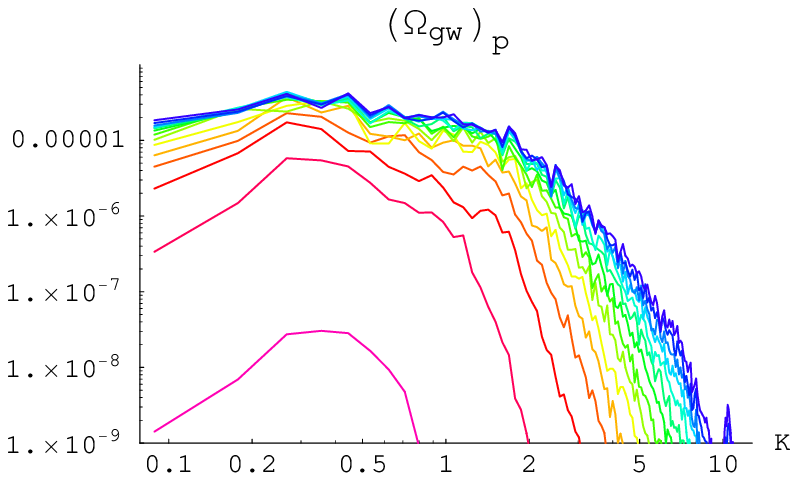}
\includegraphics[width=7cm]{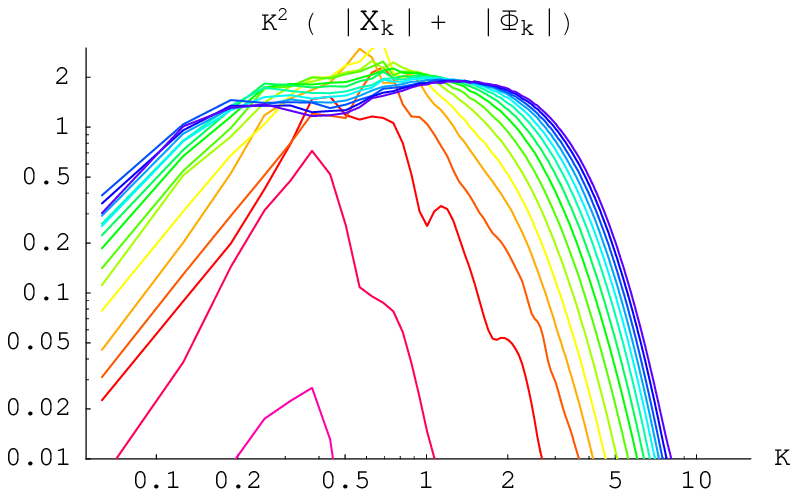}\\
\includegraphics[width=7cm]{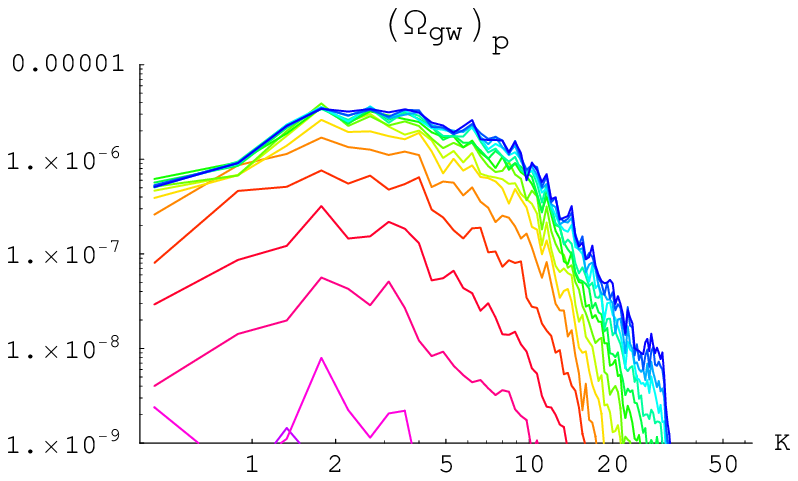}
\includegraphics[width=7cm]{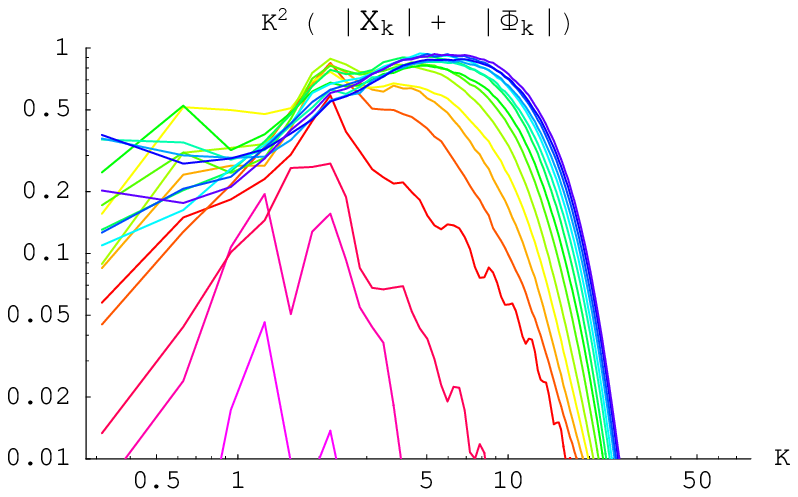}\\
\includegraphics[width=7cm]{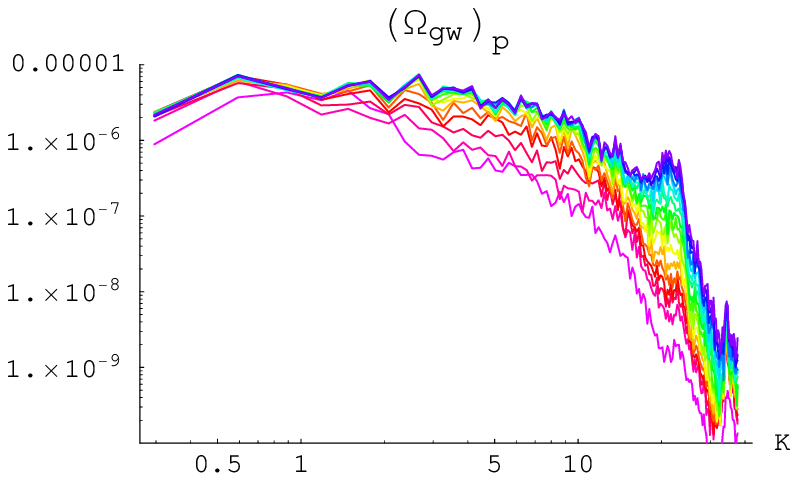}
\includegraphics[width=7cm]{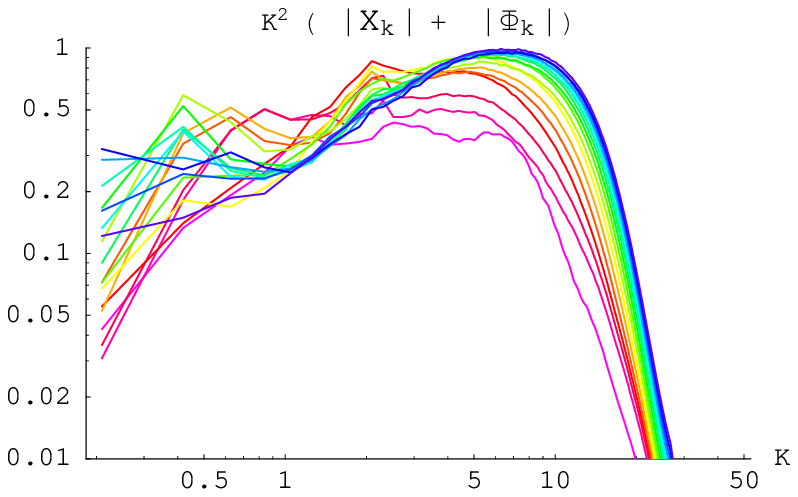}\\
\includegraphics[width=7cm]{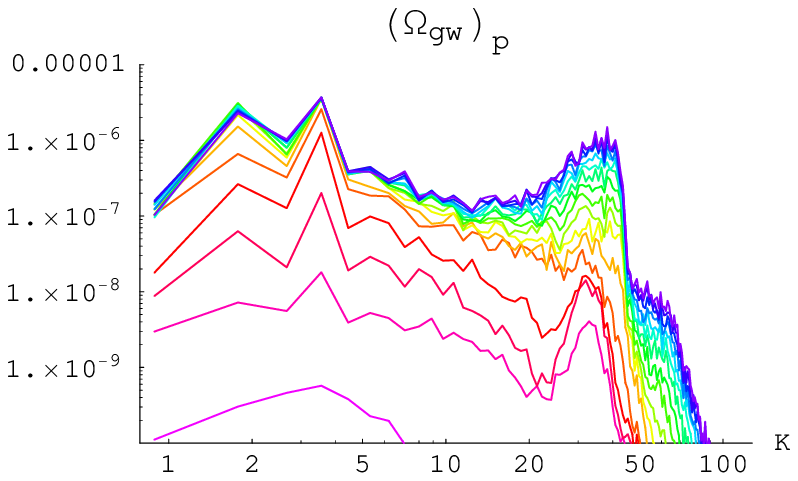}
\includegraphics[width=7cm]{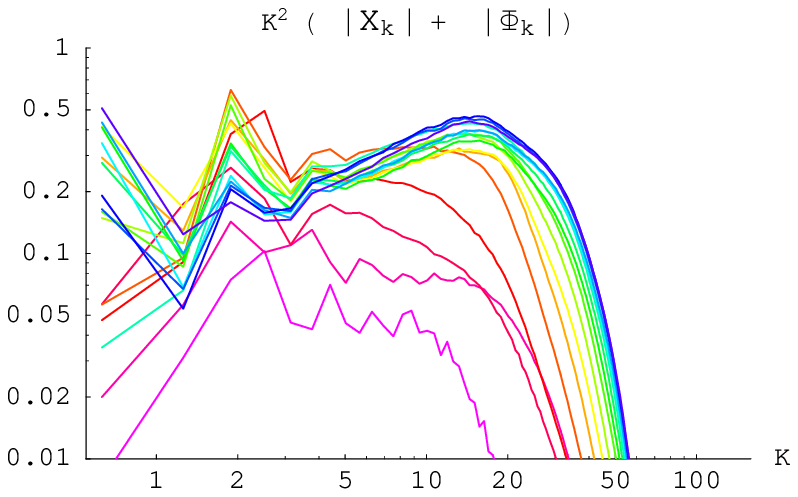}\\
\end{tabular}
\caption{Evolution with time of the gravity wave energy density spectrum (left panels) and of the scalar field spectrum (right panel) for different 
values of $q = g^2/\lambda$. From top to bottom, $q = 1.2$, $120$, $128$ and $1130$. The left panels show the spectra $\left(\Omega_{\mathrm{gw}}\right)_p$, 
see (\ref{specprod}), accumulated up to different times $x_f$, as a function of the comoving momentum $K = k/(\sqrt{\lambda} \phi_0)$. The right panels 
show the total scalar field spectrum $K^2 (|X_k| + |\Phi_k|)/\phi_0$ at different times, as a function of $K$.}
\label{diffq}
\end{center}
\end{figure}

Now we show the results of simulations for different values of the coupling constant $g^2$ and hence of the resonance parameter $q = g^2/\lambda$. 
They were all performed on lattices with $256^3$ points, up to the same final time $x_f = 240$. The spectra of gravity wave energy density 
accumulated up to different times are shown on the left panels of Fig.~\ref{diffq}, for $q = 1.2$, $120$, $128$ and $1130$. On the right 
panel we show for each of these values of $q$ the evolution with time of the spectrum of $K^2\,|X_k| + K^2\,|\Phi_k|$ for the two scalar 
fields. The first observation is that the peak amplitude in gravitational waves depends only mildly on the value of $q$ for the cases that we considered. 
In particular, we don't observe the dependence in $\Omega_{\mathrm{gw}} \propto \lambda / g^2$ derived in \cite{pregw1} and quoted in \cite{pregw2}. 
The maximal amplitude in gravity waves that we observed was for $q = 1.2$, in which case the spectrum today had a peak amplitude of 
$h^2\,\Omega_{\mathrm{gw}} \sim 5 \times 10^{-10}$ at a frequency of order $5 \times 10^6\,\mathrm{Hz}$. 

The evolution of the gravity wave spectrum is qualitatively similar to what we observed above. The IR part of the spectrum is essentially produced 
around the end of preheating, when the peak in the scalar fields spectra becomes maximal and is quickly shifted towards the UV. Later on, the UV part of 
$\Omega_{\mathrm{gw}} (k)$ increases, for instance with new peaks appearing at high frequencies in the case $q = 128$ and $q = 1130$, but the peak amplitude 
at low frequency stays basically unchanged. The frequency of this peak of the gravity wave spectrum 
depends directly on the resonant momentum $k_*$ amplified by preheating (corresponding to the initial 
peak in the scalar fields' spectra), instead for instance of the frequency of the scalar fields' modes 
($\omega_k \propto \sqrt{q}$). The corresponding frequency of the gravity waves spectrum's peak today is given by (\ref{specphi4}) 
\be
\label{fpeakphi4}
f_* \; \simeq \; \frac{k_*}{a_j\,\rho_j^{1/4}} \; 4 \times 10^{10}\,\mathrm{Hz} \; \simeq \; 
K_*\,\lambda^{1/4}\,5 \times 10^{10}\,\mathrm{Hz} \; \simeq \; K_*\,2 \times 10^{7}\,\mathrm{Hz}
\ee
where we have used $a_j^4\,\rho_j \simeq 1.15\,\lambda\,\phi_0^4 / 4$ and $k_* = \sqrt{\lambda}\,\phi_0\,K_*$. We took $\lambda = 10^{-14}$ 
is the last equality. The typical momenta $K_*$ amplified by parametric resonance tend to increase as $q^{1/4}$, see 
(\ref{kstar}), but their precise value is a non-monotonic function of $q$. For instance, for $q = 128$ (third line from the top in 
Fig.~(\ref{diffq})), we have $g^2 / \lambda = 2\,l^2$ with $l$ an integer, and in this case $k_*$ is shifted towards zero~\cite{greene}. 
Accordingly, the maximal amplitude of the gravity wave spectrum in the IR is shifted towards smaller frequencies compared to the case 
$q = 120$. For $q = 128$, the resonance is also more efficient. As a consequence, preheating ends earlier and the simulation extends more 
into the turbulent regime. 

In configuration space~\cite{bubbles}, the profile of the field inhomogeneities of $\chi$ and $\phi$ depends directly on the peak momentum 
and width of the spectra $k^2\,|\chi_k|$ and $k^2\,|\phi_k|$. During preheating, the exponential growth of the scalar fields' spectra with a 
sharp peak at the (comoving) resonant momentum $k_*$ corresponds to the amplification of bubble-like fields inhomogeneities with a characteristic 
(physical) size $R_*  \sim a / k_*$. At the end of preheating, the broadening of the scalar fields' spectra and the shift of their peaks towards the 
UV corresponds to the fragmentation of these spatial fields' fluctutations into smaller structures (see sub-section \ref{second}). 
Eq.~(\ref{bubblescoll}) estimates the contribution of this bubbly stage to the energy density in gravity waves at the time of production, 
i.e. shortly after the end of preheating. At that time, the total energy density in gravity waves is given by the peak of the spectrum, 
so this gives 
\be
\label{bubblesbis}
\left(\Omega_{\mathrm{gw}}^*\right)_p \; \sim \; \alpha\,\left(\frac{a H}{k_*}\right)_p^2 \; \sim \;
\frac{\alpha}{a_p^2\,K_*^2}\,\frac{\left(a^2\,H\right)_p^2}{\lambda\,\phi_0^2} 
\ee
where we took $R_* = a/k_*$ and $a_p$ is the value of the scale factor when the peak of the spectrum is reached.  
We factored out the product $a^2 H$ because it is constant at that time and had the same value for the different values of $q$. 
In all our simulations, the last factor in Eq.~(\ref{bubblesbis}) was: $\left(a^2\,H\right)_p^2 / \lambda\,\phi_0^2 \simeq 0.28$. 
We checked Eqs.~(\ref{bubblescoll}), (\ref{bubblesbis}) with our numerical results.  
The dependence on $1 / a_p^2$ is expected on general grounds because, 
for the model (\ref{lambdaphi4}), 
$T_{ij}^{\mathrm{TT}}$ in (\ref{Xk}) dilutes as $1/a^2$ with the expansion, 
giving an overall factor of $1/a^2$ in $S_k$. The peak amplitude of $\Omega_{\mathrm{gw}}$ is reached when the rescaled scalar modes $X_k$ and $\Phi_k$ 
stabilize to their maximal value. Gravity waves produced after that time are further suppressed by the expansion. We also observed the dependence in 
$1/k_*^2$ to the extent that we could measure this with our limited number of simulations. The value of 
$\left(\Omega_{\mathrm{gw}}^*\right)_p$ tends to decrease with $q$, for instance from about $3 \times 10^{-5}$ for $q = 1.2$ to
$3 \times 10^{-6}$ for $q = 120$. However, its precise value varies with $q$ in a non-monotomic 
way, being for instance higher for $q = 128$ than for $q = 120$. This is in agreement with the fact that the resonant momentum $k_*$ is lower for $q = 128$. 
In that case, preheating is also more rapid so that $a_p$ is lower. All in all, from our numerical results we find that Eqs.~(\ref{bubblescoll}) and 
(\ref{bubblesbis}) are satisfied with $\alpha \sim 0.15$. The precise value may not be significant - because we considered only a limited range of values 
of $q$ and because it is difficult to determine precisely the value of the scale factor ``at the time when the peak is reached'' - but the order of magnitude 
is correct. 

To summarize, for the model (\ref{lambdaphi4}), the numerical results for the frequency and the amplitude of the gravity waves spectrum are well 
described by Eqs.~(\ref{fpeakphi4}, \ref{bubblesbis}). They depend very mildly on the details of the rescattering stage, but mainly on the 
characteristic momentum $k_*$ amplified during preheating. The precise value of $k_*$ is a non-monotomic function of the parameters, but it 
may be calculated analytically for any values of the coupling constants.

\subsection{Analytical Check for Parametric Resonance}
\label{analytics}

We now study in more detail the production of gravitational radiation from scalar fields experiencing parametric resonance with large 
resonance parameter, $q >> 1$. We consider the linear stage of preheating, when only one field has been significantly amplified, so that 
Eq.(\ref{avSkpre}) is applicable. As we saw in the previous section, the main part of the gravity wave spectrum is generated during the 
subsequent evolution. However, during the linear stage, analytical solutions for the modes may be derived, and this will allow us to 
check our numerical results in details. 

\begin{figure}[hbt]
\begin{center}
\begin{tabular}{cc}
\includegraphics[width=9cm]{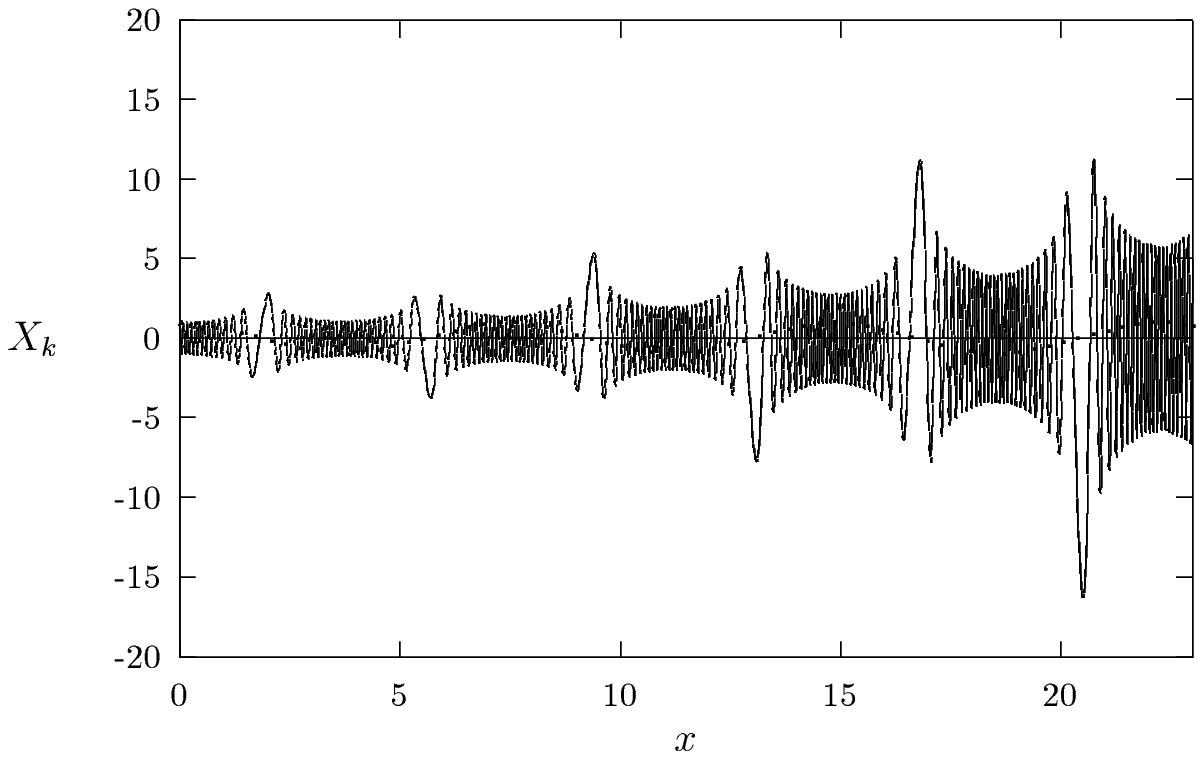}
\includegraphics[width=9cm]{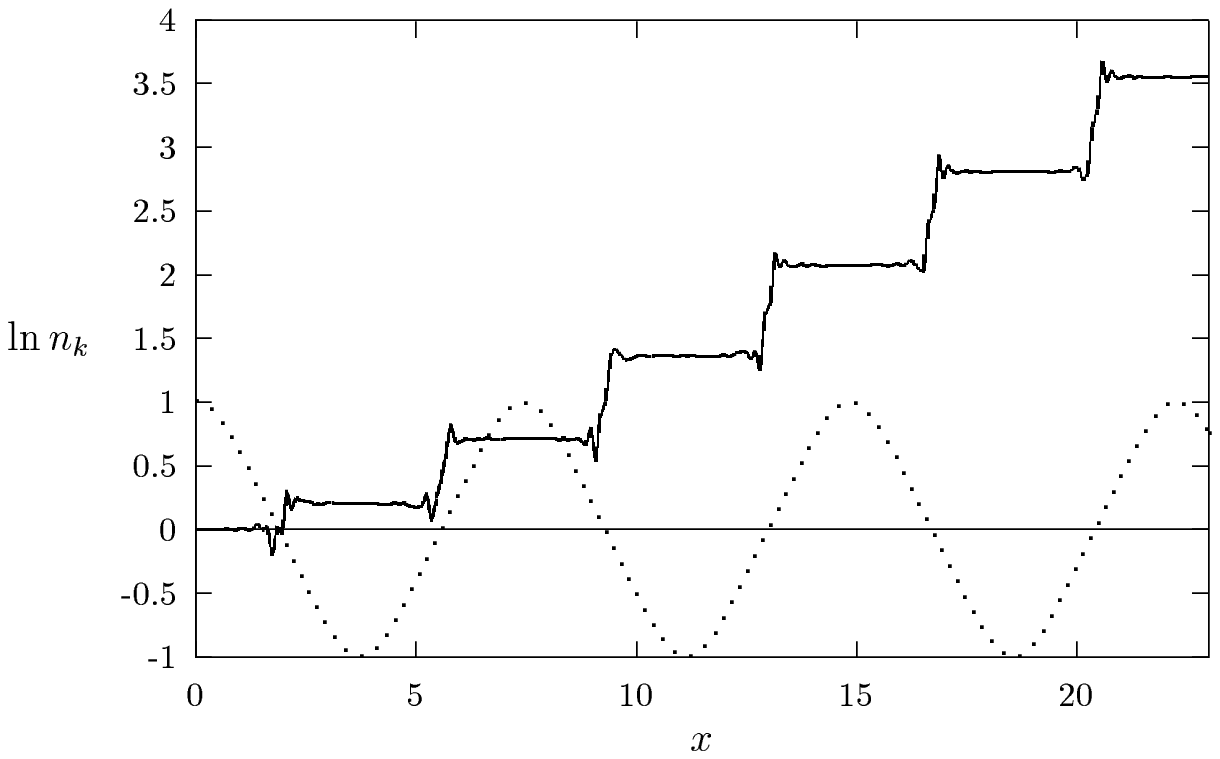}
\end{tabular}
\caption{Typical behavior of the modes amplified by parametric resonance. The left figure shows the evolution with time of the real 
part of $X_k(x)$ (see (\ref{redef})) for $k$ inside a resonance band, here for $q = 5050$. The right figure shows the evolution of the 
corresponding occupation number, together with the inflaton zero-mode. The 2 figures are taken from \cite{greene}.} 
\label{moderes}
\end{center}
\end{figure}

The energy density in gravity waves (\ref{avSkpre}) involves time integrals of the form
\be
\label{timeint}
I \, \equiv \, \int_{\tau_i}^{\tau} d\tau' \, \cos\left(k\,\tau' \right) \, a(\tau') \, \chi_p(\tau') \, \chi_{|\mathbf{k} - \mathbf{p}|}(\tau') 
\, = \, \int_{x_i}^{x} \frac{dx'}{\sqrt{\lambda} \phi_0}\,\frac{\cos[K x']}{a(x')}\, X_p(x') \, X_{|\mathbf{k} - \mathbf{p}|}(x')
\ee
where we have performed the rescalings (\ref{redef}), (\ref{Kandq}). 
The basic observation is that, during preheating, gravitational waves are produced in a step-like manner, in short intervals of time only, 
see Fig.~\ref{rhogwtime}. These correspond to the moments when the inflaton condensate crosses zero and the adiabaticity condition for 
the scalar modes is violated. The typical evolution of the modes with time is shown in Fig.~\ref{moderes}. Away from the zeros of the 
inflaton, the modes oscillate with large frequency ($\omega_k \sim \sqrt{q}$), and their amplitude evolves adiabatically. In contrast, 
their amplitude increases sharply when the inflaton vanishes (for the modes belonging to the resonance bands). Only 
these short intervals of time contribute significantly to the time integral in (\ref{timeint}). Indeed, according to the no-go theorem of 
sub-section (\ref{nogw}), no gravitational radiation is emitted by a bath of scalar modes whose frequencies evolve adiabatically with time. 
The discussion is more involved in the case of parametric resonance, because the adiabatic regime then lasts only for a limited period of time, 
between two zeros of the inflaton~\footnote{The situation here depends on the particular channel of interaction considered, namely on the 
relative signs in the phase of (\ref{staphase}), as in the ``particle-like'' interpretation of rescattering discussed in \cite{KLS}. In the 
analog of (\ref{staphase}), the time integral is now performed on a finite interval between 2 zeros of the inflaton. Some terms involve 
a phase of the form $\theta(x) = \int^x dx'\,\omega_p(x') + \int^x dx'\,\omega_{\mathbf{k} - \mathbf{p}}(x') - K x$. Because 
$\omega_p \sim \sqrt{q}$, the phase then oscillates very quickly and the time integral reduces to its contribution where $\theta(x)$ is 
stationary. This gives the condition of energy conservation for the annihilation of two $\chi$-particles into one graviton, which is not 
possible. On the other hand, when the terms in $\omega_p(x')$ and $\omega_{\mathbf{k} - \mathbf{p}}(x')$ in the phase $\theta(x)$ have 
opposite signs, the large terms in $\sqrt{q}$ cancel each other and there is no direct analogy with energy conservation between interacting
particles (the stationary phase approximation is then applicable only for large $K$ compared to half the period of the inflaton oscillations). 
In this case, the interpretation in terms of interacting wave packets is more appropriate.}. But we will confirm below (see Appendix) that 
it is sufficient to consider the vicinity of the moments of time when the inflaton vanishes.  

Consider the evolution of the modes around the $j^{\mathrm{th}}$ zero of the inflaton, at $x = x_j$. They correspond to waves scattered 
on a parabolic potential \cite{KLS}, \cite{greene}. Around its zeros (for $|x - x_j| < 1$), the inflaton is approximately linear, 
$\bar{\Phi} \propto (x-x_j)$, and Eq.~(\ref{modesequ}) reduces to
\be
\frac{d^2 X_p}{dx^2} + \left( P^2 + \frac{q}{2}\,(x - x_j)^2 \right)\,X_p = 0
\ee
where $P$ is defined in terms of the comoving momentum $p$ as in (\ref{Kandq}). The solution is given in terms of parabolic cylinder functions $W$
\be
\label{solW}
X_p(x) = A_p^j\;W\left[- \frac{P^2}{\sqrt{2q}}\, , \, (2 q)^{1/4}\,(x - x_j)\right] + 
B_p^j\;W\left[- \frac{P^2}{\sqrt{2q}}\, , \, - (2 q)^{1/4}\,(x - x_j)\right]
\ee
where $A_p^j$ and $B_p^j$ are constant coefficients. For $x < x_j$ in the adiabatic regime (for $|x - x_j| > q^{-1/4}$), Eq.~(\ref{modesequ}) 
may be solved by the WKB approximation 
\be
\label{WKB}
X_p(x) = \frac{\alpha_p^j}{\sqrt{2 \omega_p(x)}}\,e^{-i \int_{x_0}^{x_j} \omega_p(x')\,dx'} + 
\frac{\beta_p^j}{\sqrt{2 \omega_p(x)}}\,e^{i \int_{x_0}^{x_j} \omega_p(x')\,dx'}
\ee
where $\alpha_p^j$ and $\beta_p^j$ are constant (Bogoliubov-like) coefficients. 

In the Appendix we show how to match both regimes (\ref{solW}) and (\ref{WKB}). 
This allows us to follow the whole evolution of the modes with time, and therefore to calculate the integral (\ref{timeint}) 
involved in the gravity wave energy density. The result for the gravity wave spectrum (\ref{defSk}), (\ref{avSkpre}) after 
the $j^{\mathrm{th}}$ zero of the inflaton is then given by
\be
\label{Skj+1}
S_k^{j+1} = \frac{(\sqrt{\lambda} \phi_0)^6}{a^2(x_j)}\,\frac{\Mp^{-2}}{\pi^3 q}\,K^3\,\int_{-1}^1 du\,(1-u^2)^2\,\int dP\,P^6\,
\xi_p\,\xi_{|\mathbf{k} - \mathbf{p}|}\,n_p^{j+1}\,n_{|\mathbf{k} - \mathbf{p}|}^{j+1}\,\left(\mathrm{Ic}^2 + \mathrm{Is}^2\right)
\ee
where $n_p^{j+1}$ are the occupation numbers for $\chi$ in the adiabatic regime after the $j^{\mathrm{th}}$ zero of the inflaton, 
$\xi_p$ is defined in (\ref{xip}), $\mathrm{Ic}$ is defined in (\ref{Ic}) and $\mathrm{Is}$ is defined as in (\ref{Ic}) with the 
cosine replaced by a sine. Here $u = \cos( \mathbf{\hat{k}}, \mathbf{\hat{p}} )$, so that $|\mathbf{k} - \mathbf{p}|^2 = k^2 + p^2 - 2 k p u$. 

\begin{figure}[hbt]
\begin{center}
\includegraphics[width=10cm]{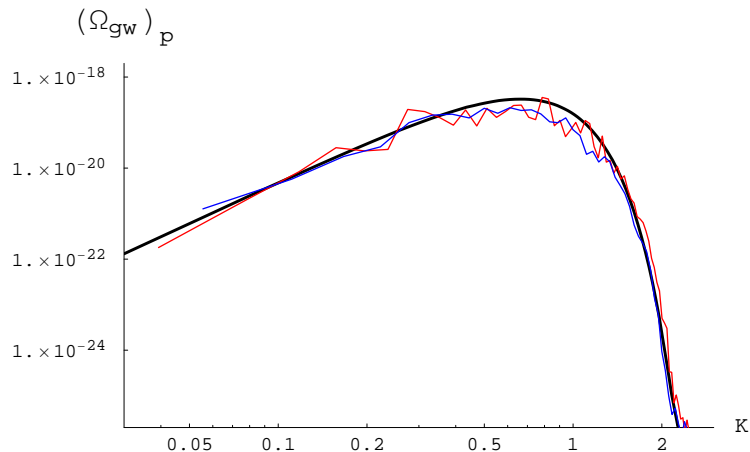}\\
\caption{Spectrum of energy density in gravity waves at the time of their production, (\ref{specprod}), 
as a function of their comoving wave-number $k$ (in units of 
$\sqrt{\lambda} \phi_0$). The $2$ thin lines correspond to the spectrum averaged over different directions in $\mathbf{k}$-space, as obtained from the 
lattice calculation. The thick line corresponds to the spectrum obtained from the analytical formula (\ref{Skj+1}). The spectra are calculated in the model 
(\ref{lambdaphi4}) with $q = 128$ during the linear stage of
preheating ($x_f \simeq 50$ in (\ref{Xk})).}
\label{check}
\end{center}
\end{figure}

We used Eq.(\ref{Skj+1}) to test our lattice calculation of the gravity wave spectrum. 
In principle, it is possible to use analytical expressions for the occupation numbers given in \cite{greene}, 
but it would be difficult to perform the integrals over $u$ and $P$ analytically. Instead, we take an interpolation 
of the occupation numbers output by the lattice simulation, and perform the integrals in (\ref{Skj+1}) 
numerically. We then compare the result to the lattice calculation in the linear stage of preheating, when only the $\chi$-field 
has been significantly amplified. Fig.~\ref{check} shows the resulting spectrum of gravity wave energy density. We see that the lattice 
calculation reproduces Eq.~(\ref{Skj+1}) very accurately. The spectra resulting from the lattice calculation, which correspond to a particular 
realisation of the initial quantum fluctuations for the scalar field, oscillate around the ensemble averaged spectrum obtained from 
Eq.~(\ref{Skj+1}). 

In Eq.~(\ref{Skj+1}), the $\mathbf{p}$-dependance comes essentially from the factor in $P^6\,n_p\,n_{|\mathbf{k} - \mathbf{p}|}$, 
which is maximal for $p \simeq |\mathbf{k} - \mathbf{p}| \simeq p_*$. For $k << p_*$, the two integrals in (\ref{Skj+1}) do not depend 
on $k$. For $K/\sqrt{2q} < \mathrm{few} \times 2\pi$, the cosine in (\ref{Ic}) may go out of the time integral, and it cancels with the 
corresponding sine coming from $\mathrm{Is}$. Therefore, at small wave-numbers, the only $k$-dependence in Eq.~(\ref{Skj+1}) comes from 
the factor $K^3$ in front of the integrals, which determines the infrared slope of the spectrum in Fig.~\ref{check}.

\section{Expectations for Preheating after Hybrid Inflation}
\label{hybrid}

In this section, we provide preliminary estimates for the gravity 
waves emitted from preheating after hybrid inflation. The mechanism 
for preheating is different than for the model considered in the 
previous section, but the resulting fragmentation of bubble-like 
field inhomogeneities is qualitatively similar. We will therefore 
extrapolate to this case the results obtained before for the 
peak frequency and amplitude of the gravity wave spectrum emitted 
from the bubbly stage. We will analyze this case in greater detail in
a subsequent publication \cite{gwhyb}.

In the model of chaotic inflation that we considered in the previous 
section, the energy density during preheating is too high for the 
resulting gravity waves to fall into the frequency range accessible 
to interferometric direct detection experiments (from 
$f \sim 10^{-4}\,\mathrm{Hz}$ corresponding to the lower limit of LISA, 
to $f \sim 10^{3}\,\mathrm{Hz}$ corresponding to the upper limit of 
LIGO). Indeed, assuming radiation domination at the time of production, 
the gravity wave frequency today (45) may be re-written
\be
\label{fHp}
f \simeq \frac{k_p}{H_p}\,\frac{\rho_p^{1/4}}{10^8\,\mathrm{GeV}}
\,1\,\mathrm{Hz}
\ee 
where $H_p$ and $\rho_p$ are the Hubble rate and the total energy 
density at the time of gravity waves emission, and $k_p$ is their 
physical wave-number at that time. Since we can only expect a significant 
amplitude in gravity waves for wavelengths inside the Hubble radius, 
$k_p > H_p$, it is clear from Eq. (\ref{fHp}) that models with low 
energy scales are more interesting from an observational perspective. 
Note however that the peak of the gravity waves spectrum 
generally occurs at higher frequencies, $k_p >> H_p$. 

In hybrid models, the inflationary energy scale is typically a free 
parameter ; it could just as easily be of the order of the GUT scale 
or of the order of electroweak scale. At the end of hybrid inflation, 
the inflaton decays via a spinodal instability of the inhomogeneous 
modes accompanying the symmetry breaking, in the process of tachyonic 
preheating \cite{tach1, tach2, tach3}. This also involves 
inhomogeneous spatial structures, whose typical size 
$R_*$ depends on the model considered. Based on Eqs.~(\ref{fpeakphi4}) 
and (\ref{bubblesbis}), we estimate the present-day frequency 
and amplitude associated with the peak of the gravity wave spectrum 
emitted from this bubbly stage as 
\be
\label{fstar}
f_* \sim \frac{4 \times 10^{10}\,\mathrm{Hz}}{R_*\,\rho_p^{1/4}}
\ee
\be
\label{omegastar}
h^2\,\Omega_{\mathrm{gw}}^* \sim 10^{-6}\,\left(R_*\,H_p\right)^2
\ee

Hybrid inflation should generally end quickly compared to the 
expansion rate of the universe, the so-called waterfall constraint. 
It follows that for practical purposes, one may usually neglect the 
expansion of the universe during preheating. The energy density 
$\rho_p$ above is then just given by the energy scale during 
inflation, $\rho_p = V_{\mathrm{inf}}$, and $H_p$ is the corresponding 
Hubble rate. Note that the waterfall constraint implies typically 
$R_* > 1 / H_p$, so this already places some lower limit on $f_*$ 
and some upper limit on $\Omega_{\mathrm{gw}}^*$. 

We will estimate (\ref{fstar}) and (\ref{omegastar}) as a function 
of the parameters for two particular models of tachyonic preheating.    
A prototype of hybrid inflation model is given by the potential
\be
\label{pothyb}
V = \frac{\lambda}{4}\,\left(\sigma^2 - v^2\right) + 
\frac{g^2}{2}\,\phi^2\,\sigma^2 + \frac{m^2}{2}\,\phi^2
\ee
For $\phi > \phi_c = \frac{\sqrt{\lambda}}{g}\,v$, the fields have 
positive effective mass squared, and the potential has a valley at 
$\sigma = 0$. Inflation occurs while $\phi$ decreases slowly in this 
valley. The energy scale during inflation is usually dominated by the 
false vacuum contribution, $V_{\mathrm{inf}} = \frac{\lambda}{4}\,v^4$. 
When $\phi$ reaches the bifurcation point $\phi_c$, inflation ends 
and the fields roll rapidly towards the global minimum at $\phi = 0$ 
and $|\sigma| = v$. To avoid the production of domain walls, one 
usually considers a complex field $\sigma$. 

When $\phi$ reaches $\phi_c$, the curvature of the effective potential 
becomes negative and field inhomogeneities are amplified by the tachyonic 
instablilty. The details of this process, in particular the typical momenta 
$k_*$ that are amplified, depend on the particular trajectory followed by the 
fields $\phi$ and $\sigma$, which in turn depends on the values of the 
coupling constants and on the initial inflaton velocity at $\phi = \phi_c$. 
Here we will only consider two particular trajectories in field space, 
which lead to different predictions for the gravity waves spectrum. 
Preheating in these two cases has been studied in detail in 
\cite{tach1}, \cite{tach2}, and we will rely on their results for 
our estimations.

For $g^2 >> \lambda$, or sufficient inflaton velocity at $\phi = \phi_c$, 
the inflaton overshoots the bifurcation point and the field $\sigma$ falls 
down only when when $\phi << \phi_c$. In this case, the effective potential 
for $\sigma$ is dominated by~\cite{tach1}
\be
\label{-m2sigma2}
V = \frac{\lambda}{4}\,v^4 - \frac{\lambda}{2}\,v^2\,\sigma^2 
+ \frac{\lambda}{4}\,\sigma^4
\ee
The typical momenta amplified by preheating in this model vary as 
$\sqrt{\lambda} v / \sqrt{\ln\left(\frac{100}{\lambda}\right)}$. From the 
numerial results presented in \cite{tach2}, we estimate the typical 
size of the inhomogeneous structures of $\sigma$ when they fragment as
\be
R_* \sim \frac{5}{\sqrt{\lambda}\,v}\,
\sqrt{\ln\left(\frac{100}{\lambda}\right)}
\ee
The resulting frequency (\ref{fstar}) and amplitude 
(\ref{omegastar}) associated with the peak of the gravity wave spectrum 
are then given by
\be
f_* \sim \frac{\lambda^{1/4}\,10^{10}\,\mathrm{Hz}}
{\sqrt{\ln\left(\frac{100}{\lambda}\right)}}
\ee
and
\be
h^2\,\Omega_{\mathrm{gw}}^* \sim 2 \times 10^{-6}\,
\ln\left(\frac{100}{\lambda}\right)\,\left(\frac{v}{\Mp}\right)^2
\ee
Note that $f_*$ depends on the energy scale only through the coupling 
constant $\lambda$, and that this dependence is very weak, essentially 
varying as $\lambda^{1/4}$. The peak of the gravity wave spectrum may fall 
into the range accessible to interferometric direct detection experiments, 
$f_* < 10^2\,\mathrm{Hz}$, only for extremely small values of the 
coupling constant, $\lambda < 10^{-28}$. 

Another special case corresponds to $g^2 = 2\,\lambda$ in (\ref{pothyb}) and 
sufficiently low inflaton velocity at the critical point $\phi = \phi_c$. 
In this case, the effective potential has the same structure as the effective 
potential of SUSY-inspired F-term inflation, and the fields $\phi$ and $\sigma$ 
fall down along a simple linear trajectory, see e.g.~\cite{levtalk}. 
Along this trajectory, the effective potential for the field $\sigma$ takes 
the form (up to the redefinition $\lambda \rightarrow \lambda / 3$)
\be
V = \frac{\lambda}{12}\,v^4 - \frac{\lambda}{3}\,v\,\sigma^3 
+ \frac{\lambda}{4}\,\sigma^4
\ee 
Preheating in this model has been investigated in detail in \cite{tach2}. 
The typical momenta amplified vary as $k_* \sim \frac{\lambda v}{2\pi}$. 
The process is initially dominated by fluctuations with momenta somewhat 
greater than $k_*$, whose initial amplitude is more suppressed, but which 
grow faster. It follows that the bubble-like field inhomogeneities
initially have a typical size $R_*$ somewhat smaller than $1/k_*$, but they are 
also quite distant from each other and they expand for some time before 
colliding. We estimate the typical size of these spatial structures when 
they collide from the numerical results of \cite{tach2}
\be
R_* \sim \frac{5}{\lambda\,v}
\ee
The resulting frequency (\ref{fstar}) and amplitude 
(\ref{omegastar}) associated with the peak of the gravity wave spectrum 
are then given by
\be
\label{fcubic}
f_* \sim \lambda^{3/4}\,10^{10}\,\mathrm{Hz}
\ee
and
\be
\label{omegacubic}
h^2\,\Omega_{\mathrm{gw}}^* \sim \frac{2 \times 10^{-6}}{\lambda}\,
\left(\frac{v}{\Mp}\right)^2
\ee
Note that this case generally leads to higher amplitude and smaller 
frequency in gravity waves than the previous model (\ref{-m2sigma2}). 
Requiring the frequency of the peak of the gravity wave spectrum to 
fall into the range accessible to interferometric direct detection 
experiments, $f_* < 10^2\,\mathrm{Hz}$, gives $\lambda < 2 \times 10^{-11}$. 
This is still a very small value for the coupling constant, but it is 
considerably higher than for the model (\ref{-m2sigma2}). 

For example, according to Eqs. (\ref{fcubic}) and (\ref{omegacubic}), 
a model with $g^2 = 2\,\lambda = 2 \times 10^{-11}$ and $v = 10^{12}\,\mathrm{GeV}$ 
(so that the waterfall constraint is marginally satisfied) would lead 
to $f_* \sim 50\,\mathrm{Hz}$ and 
$h^2\,\Omega_{\mathrm{gw}}^* \sim 2 \times 10^{-9}$, which could already be 
detectable by LIGO. Note that we only considered here the peak of the 
gravity wave spectrum, but the infrared tail of the spectrum could also 
be observationally relevant.  

The preliminary estimates made in this section are very simplified and 
should be supplemented by numerical simulations~\cite{gwhyb}. 
We considered only the peak of the gravity wave spectrum emitted during the 
bubbly stage for two particular models, neglecting the amplification 
of the field $\phi$ or any degree of freedom other than $\sigma$. 
In particular, other models should be considered, and a significant amount 
of gravity waves may still be produced during the transition from the 
bubbly stage to the regime of well-developed turbulence\footnote{In the 
chaotic inflation model studied in the previous section, the energy density 
in gravity waves produced during this transition stage is suppressed by the 
expansion of the universe, but we don't expect such a suppression for hybrid 
inflation models.}. However, we see that even in simple models, the results 
for the gravity wave spectrum may cover a wide range of values.

\section{Discussion and Summary}
\label{conclusions}

In this paper we constructed a general formalism to calculate the 
spectrum of classical gravitational waves emitted from random media
of dynamical scalar fields in an expanding universe. In principle 
this can be applied to any cosmological situation where
scalar field dynamics is important and potentially interesting for the 
production of gravitational radiation. This includes, for instance, 
early universe phase transitions at the electroweak energy scale or 
above \cite{Grojean:2006bp}, the decay of the inflaton field in the process
of (p)reheating after inflation, and other non-equilibrium processes.

In the paper we advanced the theoretical framework for preheating after 
inflation, which is accompanied by the violent decay of the inflaton and 
the copious production of bosonic fields coupled to it. 
Based on our formalism, we have developed a numerical code to calculate gravity 
wave production. 

The formalism and numerics for gravity wave emission in this paper were 
implemented for models of chaotic inflation, and in particular for chaotic inflation 
with a quartic potential. This case served as a convenient ground to contrast 
our numerical results with previous ones, and to test them analytically 
in great detail. The dominant contribution to gravity wave emission comes 
from the non-linear ``bubbly'' stage at the end of preheating. Our numerical 
results for the present-day frequency and amplitude of the resulting 
spectrum's peak are well described by the simple formulas
\be
\label{conclu}
f_* \sim \frac{4 \times 10^{10}\,\mathrm{Hz}}{R_*\,\rho_p^{1/4}} \;\;\; \mathrm{ and }  
\;\;\; h^2\,\Omega_{\mathrm{gw}}^* \sim 10^{-6}\,\left(R_*\,H_p\right)^2
\ee
where $H_p$ and $\rho_p$ are the Hubble rate and total energy density during 
the bubbly stage. The characteristic (physical) size $R_* \sim a / k_*$ of the 
bubble-like field inhomogeneities depends on the typical (comoving) momentum 
$k_*$ amplified by preheating. Up to factors depending on the expansion of 
the universe, $f_*$ and $h^2\,\Omega_{\mathrm{gw}}^*$ are then known 
functions of the parameters for different models. The present day frequency 
$f_*$ depends on the particular chaotic model considered, but it remains too high, 
above $10^6$ Hz, to be detected by currently planned experiments. The fraction 
of energy density in gravity waves also varies from one chaotic model to 
another. It is generally lower for the model $m^2 \phi^2 +g^2 \phi^2 \chi^2$ than 
for the model with $\lambda\,\phi^4$ inflation, because the typical momenta 
amplified $k_* \sim \sqrt{g m \Phi_0}$ are higher in that case. The results for  
$\rho_{\mathrm{gw}}/\rho_{\mathrm{tot}}$ may also be different in models with 
trilinear or non-renormalisable interactions \cite{tri}, where the dynamics 
ocurs faster so that $(R_*\,H)$ is less diluted by the expansion of the universe.

Based on Eq.~(\ref{conclu}), we also estimated the frequency and the amplitude 
of the peak of the gravity-wave spectrum produced from the bubbly stage in two 
particular models of preheating after hybrid inflation. Because there are more 
free parameters in this case, the results may cover a much wider range of values.  
We found that in principle, for models with very small coupling constants, the 
resulting gravity waves spectrum may be relevant already for LIGO/VIRGO. 
Gravity waves from preheating after hybrid inflation certainly deserve further 
investigations. 

Besides gravity wave interferometers probing typical frequencies which range
from $10^{-4}$ Hz (LISA) up to $10^3$ Hz (LIGO), there are several bar or spherical 
resonant detectors (see e.g.~\cite{minigrail}) operating in the kHz range. Experiments 
have also been proposed at higher frequencies, up to $100$ Mhz~\cite{Nishizawa:2007tn}. 
However, the sensitivity to $h^2\,\Omega_{\mathrm{gw}}$ drops when the frequency 
increases, and the gravity waves produced from preheating after chaotic inflation lie 
outside the range currently accessible to these experiments.    

The frequencies of gravitational waves emitted from preheating could 
be naturally redshifted to lower values if there was some intermediate 
matter dominated stage between preheating and the radiation dominated 
era. If the scale factor expands by a factor $a_m$ during this stage, 
then the frequencies of gravity waves emitted before that stage are 
decreased by a factor of $1/a_m^{1/4}$. However, the fraction of 
energy density in gravitational radiation is then diluted by a factor 
of $1/a_m$. 

We only considered the dynamics of two coupled fields during preheating, 
the inflaton and another scalar field. Clearly, the production of gravity 
waves will increase with the number of degrees of freedom. 
In particular, non-scalar bosonic degrees of freedom 
excited during preheating will change the results for gravity wave 
production during the turbulent evolution towards thermal equilibrium. 
Indeed, we found that neither established Kolmogorov 
turbulence for scalars nor scalars in thermal equilibrium contribute 
to $\rho_{\mathrm{gw}}$. The situation will be different for gauge bosons 
and photons. It will be interesting to address the issue of gravity 
waves emission from the vector fields excited during preheating.


\section{Acknowledgements}

We would like to thank Latham Boyle and Andrei Frolov for stimulating discussions, 
and Richard Easther, Daniel Figueroa, Juan Garcia-Bellido and Eugene Lim for useful
discussions and correspondences during different stages of the project.
A.B. and G.F. were supported by NSF grant PHY-0456631. L.K. was supported by NSERC and CIFAR.


\section{Appendix}
\label{appB}

In this Appendix, we derive the evolution with time of the modes $X_p$ of sub-section \ref{analytics} 
during parametric resonance in the model (\ref{lambdaphi4}). This allows us to calculate analytically 
the gravity wave energy density spectrum emitted during this stage. 

Around the zeros of the inflaton field, the modes $X_p$ are given by Eq.~(\ref{solW}). 
In between the zeros of the inflaton, they are given by the WKB solution (\ref{solW}).
In this regime, the occupation numbers are constant and well defined. 
With our definition of $\beta_p^j$ above, they are given by
\be
\label{npj}
n_p^j = \sqrt{\lambda} \phi_0\,|\beta_p^j|^2
\ee
where the prefactor comes from the fact that $\omega_k$ is the rescaled frequency, see (\ref{modesequ}) and (\ref{Kandq}). 
Note that $X_p$ and $\beta_p^j$ have dimension $\mathrm{mass}^{-1/2}$, as the modes $\chi_p$ in (\ref{aa+}), so that $n_k$ is dimensionless.

Both solutions (\ref{solW}) and (\ref{WKB}) overlap on an interval of $x$ of order $q^{-1/4} < |x - x_j| < 1$. Matching the four coefficients 
gives
\ba
\label{AetB}
B_p^j = \frac{(2 q)^{-1/8}}{\sqrt{2 \xi_p}}\;
\left[ \alpha_p^j\,e^{-i (\pi/4 + \theta_p^j - \zeta_p/2)} + \beta_p^j\,e^{i (\pi/4 + \theta_p^j - \zeta_p/2)} \right] 
\nonumber \\
A_p^j = i\,\sqrt{\frac{\xi_p}{2}}\,(2 q)^{-1/8}\;
\left[ \alpha_p^j\,e^{-i (\pi/4 + \theta_p^j - \zeta_p/2)} + \beta_p^j\,e^{i (\pi/4 + \theta_p^j - \zeta_p/2)} \right]
\ea
where 
\be
\label{xip}
\xi_p = \left( 1 + e^{-2\pi P^2/\sqrt{2q}} \right)^{1/2} - e^{-\pi P^2/\sqrt{2 q}} 
\ee
The phases $\theta_p^j$ and $\zeta_p$ are given in \cite{KLS}, \cite{greene}. Their precise value will not be important for our purpose. 
For $x > x_j$, the modes are again given by the adiabatic solution (\ref{WKB}) with new coefficients $\alpha_p^{j+1}$ and $\beta_p^{j+1}$. 
Matching this new solution with (\ref{solW}) gives the evolution of $\alpha_p$, $\beta_p$ and $n_p$ around the $j^{\mathrm{th}}$ zero of 
the inflaton. For large occupation numbers, $n_p^{j+1}$ at $x > x_j$ is related to $n_p^j$ at $x < x_j$ by~\cite{KLS}
\be
\label{npj+1}
n_p^{j+1} = \left( 1 + 2\,e^{-2\pi P^2/\sqrt{2q}} - 2\,\sin \Theta_p^j\,e^{-\pi P^2/\sqrt{2 q}}\,\sqrt{1 + e^{-2\pi P^2/\sqrt{2q}}} \right)\;n_p^j
\ee
where $\Theta_p^j = 2\,\theta_p^j - \zeta_p + \mathrm{arg} \beta_p^j - \mathrm{arg} \alpha_p^j$. 

Because the expansion of the universe may be scaled out of the equations for the model (\ref{lambdaphi4}), the resonance involves separate stability and 
instability bands for the momenta $p$, which depend on the phase $\Theta_p^j$. Only the modes belonging to the instability bands are amplified and 
contribute to the $p$-integral in the gravity wave energy density (\ref{avSkpre}). The maximal occupation number is reached for $\sin \Theta_p^j = -1$ 
in (\ref{npj+1}), corresponding to constructive interference, and it is a good approximation to take this value for all the modes under consideration. 
In this case, Eq.~(\ref{AetB}) simplifies to
\be
B_p^j = \sqrt{\frac{2}{\xi_p}}\,\frac{\beta_p^j}{(2 q)^{1/8}}\,e^{i (\pi/4 + \theta_p^j - \zeta_p/2))} \;\;\;\;\; \mbox{ and } \;\;\;\;\; A_p^j = 0
\ee
where we have used the fact that $\alpha_p^j$ and $\beta_p^j$ differ only by a phase in the limit of large occupation numbers. Finally, using (\ref{npj}) 
and (\ref{npj+1}) with $\sin \Theta_p^j = -1$, we may relate $B_p^j$ to the occupation numbers before and after $x = x_j$
\be
\label{Bpnp}
|B_p^j|^2 \, = \, (2 q)^{-1/4}\,\frac{2}{\xi_p}\,\frac{n_p^j}{\sqrt{\lambda} \phi_0} \, = \, \frac{2 \xi_p}{(2 q)^{1/4}}\,\frac{n_p^{j + 1}}{\sqrt{\lambda} \phi_0} 
\ee

We are now ready to calculate the time integral (\ref{timeint}). Since the amplitude of the modes is increasing exponentially with time, it is sufficient  
to perform the integral over the latest inflaton oscillation only. The solution (\ref{solW}) has a peak amplitude at $x = \tilde{x}_j$, just after $x = x_j$. 
It is accurate over an interval of $x$ of order $\Delta x = \mathrm{few} \times q^{-1/4}$, around $x = x_j$, and differs from the exact solution only far 
inside the adiabatic regime. The amplitude of the modes at that time is lower than their amplitude at $x = \tilde{x}_j$ by a factor of order $q^{-1/8}$, as may 
be seen from Eqs.~(\ref{solW}), (\ref{WKB}) and (\ref{AetB}). This gives a relative contribution of order $q^{-1/2}$ to the gravity wave energy density 
(\ref{avSkpre}), which is negligible for $q >> 1$. The integral (\ref{timeint}) after the $j^{\mathrm{th}}$ zero of the inflaton is therefore well approximated by  
\be
I^{j+1} \, \simeq \, \frac{B_p^j\,B_{|\mathbf{k} - \mathbf{p}|}^j}{\sqrt{\lambda} \phi_0\,a(x_j)}\;\int_{x_j - \epsilon}^{x_j + \epsilon} dx'\,\cos[K x']\;
W\left[- \frac{P^2}{\sqrt{2q}}\, , \, - (2 q)^{1/4}\,(x - x_j)\right]\;W\left[- \frac{|\mathbf{K} - \mathbf{P}|^2}{\sqrt{2q}}\, , \, - (2 q)^{1/4}\,(x - x_j)\right]
\ee
where $\epsilon = \mathrm{few} \times q^{-1/4}$. The scale factor $a(x)$ goes out of the integral since it is approximately constant on the interval 
$\left[\tilde{x}_j - \epsilon \, , \, \tilde{x}_j - \epsilon \right]$. Introducing the variable $v = (2 q)^{1/4}\,(x - x_j)$ and using (\ref{Bpnp}), we have
\be
\label{Ij+1}
\left| I^{j+1} \right|^2 = \frac{a^{-2}(x_j)}{(\sqrt{\lambda} \phi_0)^4}\;\frac{2}{q}\;
\xi_p\,\xi_{|\mathbf{k} - \mathbf{p}|}\,n_p^{j+1}\,n_{|\mathbf{k} - \mathbf{p}|}^{j+1}\,\mathrm{Ic}^2
\ee
where 
\be
\label{Ic}
\mathrm{Ic} = \int_{v_i}^{v_f} dv\,\cos\left[\frac{K\,v}{\sqrt{2q}} - K\,x_j\right]\;W\left[- \frac{P^2}{\sqrt{2q}}\, , \, - v \right]\;
W\left[- \frac{|\mathbf{K} - \mathbf{P}|^2}{\sqrt{2q}}\, , \, - v \right]
\ee
The result does not depend on the precise values of $v_i$ and $v_f$ provided the integration range $[v_i, v_f]$ extends sufficiently into the adiabatic regime. 
The functions $W$ depend only slightly on their first argument, which is smaller than one for modes inside the instability bands (see (\ref{kstar})). They are 
oscillatory for all $v$, with a peak amplitude at $v = \tilde{v} \simeq 1.5$. The amplitude of the oscillations decreases as $v$ moves away from $\tilde{v}$, 
and the oscillations at $v > \tilde{v}$ have larger amplitude than the ones at $v < \tilde{v}$. The main contribution to the integral comes from the oscillation 
centered around $\tilde{v}$, whose width is about $\Delta v \simeq 4$. For a precise calculation, we may also take into account the first oscillation before 
$\tilde{v}$ and the first two oscillations after it, corresponding to $v_i \simeq - 4$ and $v_f \simeq 7$.      

Finally, inserting (\ref{Ij+1}) into (\ref{avSkpre}) and performing the integral over the azimuthal angle of $\mathbf{p}$, we derive Eq.(\ref{Skj+1}) 
for the gravity wave energy density spectrum after the $j^{\mathrm{th}}$ zero of the inflaton.



\end{document}